\documentclass[aps,prd,preprint,superscriptaddress,showpacs]{revtex4}
\usepackage{graphicx}

\usepackage{ulem}
\usepackage{color}
\definecolor{My_red}        {cmyk}{0.00,1.00,1.00,0.20}

\newcommand{\bmat}{\left(\begin{array}}
\newcommand{\emat}{\end{array}\right)}
\newcommand{\beq}{\begin{equation}}
\newcommand{\eeq}{\end{equation}}
\newcommand{\wt}{\widetilde}

\def\ra{\rightarrow}

\def\ld{\lambda}
\def\f{\frac}

\def\bwt{\begin{widetext}}
\def\ewt{\end{widetext}}
\def\be{\begin{equation}}
\def\ee{\end{equation}}
\def\bea{\begin{eqnarray}}
\def\eea{\end{eqnarray}}
\def\bean{\begin{eqnarray*}}
\def\eean{\end{eqnarray*}}
\def\bary{\begin{array}}
\def\eary{\end{array}}
\def\bit{\begin{itemize}}
\def\eit{\end{itemize}}

\def\ra{\rightarrow}

\def\ld{\lambda}

\def\su5u1{SU(5) \times U(1)}
\def\fsu5u1{SU(5) \times U(1)'}
\def\so10{SO(10)}
\def\sq20{SO(10) \times SO(10)}

\def\ra{\rightarrow}

\def\ld{\lambda}
\def\f{\frac}

\def\L{\left(}
\def\R{\right)}

\def\ra{\rightarrow}

\def\ld{\lambda}

\def\su5u1{SU(5) \times U(1)}
\def\fsu5u1{SU(5) \times U(1)'}
\def\so10{SO(10)}
\def\sq20{SO(10) \times SO(10)}

\usepackage[centertags]{amsmath}
\usepackage{amssymb}

\begin{document}

\title{On  Naturalness of the MSSM and NMSSM}

\author{Zhaofeng Kang}
\email{zhfkang@itp.ac.cn}
\affiliation{ State Key Laboratory of Theoretical Physics,
      Institute of Theoretical Physics, Chinese Academy of Sciences,
Beijing 100190, P. R. China }

\author{Jinmian Li}
\email{	jmli@itp.ac.cn}
\affiliation{ State Key Laboratory of Theoretical Physics,
      Institute of Theoretical Physics, Chinese Academy of Sciences,
Beijing 100190, P. R. China }

\author{Tianjun Li}
\email{tli@itp.ac.cn}

\affiliation{ State Key Laboratory of Theoretical Physics,
      Institute of Theoretical Physics, Chinese Academy of Sciences,
Beijing 100190, P. R. China }

\affiliation{George P. and Cynthia W. Mitchell Institute for
Fundamental Physics, Texas A$\&$M University, College Station, TX
77843, USA }

\date{\today}

\begin{abstract}

With a bottom-up approach, we consider naturalness in the MSSM and NMSSM. Assuming the light stops,
the  LHC gluino search implies that the degree of fine tuning in both models is less than
$2.5\%$. Taking the LHC hints for the SM-like Higgs boson mass $m_h\sim125$ GeV seriously, 
we find that naturalness will favor the NMSSM. We study the Higgs boson mass for
several scenarios in the NMSSM: (1) A large $\ld$ and the doublet-singlet Higgs boson
mixing effect pushing upward or pulling downward $m_h$. The 
former case can readily give the di-photon excess of the  Higgs boson decay 
whereas the latter case can not. However, we point out that the former case has a new large fine-tuning related to
strong $\ld-$RGE running effect and vacuum stability.
(2) A  small $\ld$ and the mixing effect pushing $m_h$ upward. Naturalness status
becomes worse and no significant di-photon excess can be obtained. In these scenarios,
the lightest supersymmetric particle (LSP) as a dark matter candidate is strongly
disfavored by the XENON100 experiment. Even if the LSP can be a viable dark matter
candidate, there does exist fine-tuning. The above naturalness evaluation is based
on a high mediation scale for supersymmetry breaking, whereas for a low mediation
scale, fine-tuning can be improved by about one order.

\end{abstract}


\maketitle

\section{Introduction and motivations}

Supersymmetry (SUSY) solves the gauge hiearchy problem in the Standard Model (SM) naturally.
In the supersymmetric SMs with $R$-parity, gauge coupling unification can be achieved,
which strongly indicates the Grand Unified Theory (GUT). Also,
a surprising gift is that the lightest supersymmetric particle (LSP) services as
a cold dark matter (DM)  candidate. The recent LHC experiments and DM search experiments
lead to the following implications:

\begin{itemize}
  \item The CMS~\cite{CMS:SUSY} and ATALS~\cite{ATLAS:SUSY} SUSY searches constrain
the colored particles stringently and push their masses towards the TeV region, except that
the light stops  $\sim{\cal O}(200)$ GeV can still be allowed~\cite{lightstop, Brust:2011tb}.
Light stops are good for naturalness by virtue of the absence of
enormous fine-tuning from the stop radiative corrections on the Higgs soft mass square
      \begin{align}\label{mHu}
m_{H_u}^2\sim -\f{3h_t^2(m_{Q_3}^2+m_{U_3^c}^2)}{8\pi^2}\log \f{M}{m_{\rm SUSY}}\sim m_{\wt t}^2,
  \end{align}
where $M$ is the SUSY-breaking mediation scale. Moreover, they maintain the discovery
potential of colored supersymmetric particles (sparticles) at the early LHC run.

  \item In the Minimal Supersymmetric SM (MSSM), the lightest CP-even Higgs boson mass is less than
the $Z$-boson mass $m_Z$, and is lifted by the top and stop loop corrections.
However, the recent Higgs search results at the ATLAS and CMS experiments imply a SM-like Higgs
boson with mass around $\sim125$ GeV~\cite{Higgs:126}, which inspired extensive studies
within the MSSM~\cite{Higgs:CMSSM,Shih,Barger}.
Such a relatively heavy Higgs boson typically requires TeV scale stops and an anomalously large $|A_t|$.
In the gauge mediated symmetry breaking (GMSB) scenario, a large $|A_t|$ is viable only
when the messenger scale is very high~\cite{Shih}, which is a prediction
in the ``asymmetric" gravitino scenario~\cite{Kang:2011ny}.

  \item The discovery of the relatively heavy Higgs boson and the null XENON100 experimental result~\cite{XENON100}
  place the neutralino LSP DM candidate in an unsatisfying position~\cite{Barger}.
 %To make $m_h\simeq125$ GeV,
 Now the LSP abundance tends to be far
 above the WMAP-measured value except for very fine-tuned coannhilations or for a Higgsino-like LSP
 which is, however, strongly disfavored by the XENON100. Note that it can constrain on the
 $\mu$ parameter and the naturalness thereof.
%   On the other hand, the null result from the  experiment tightly constraints the LSP
%Higgsino component

%and in turn roughly gives the lower bound of the
%Thus,the XENON100 null result and the LHC discovery of Higgs boson worsens the naturalness of SUSY models.
 \end{itemize}
In summary,  the naturalness is quite  pessimistic in the mSUGRA-like MSSM,
although it is still far from the failure allegation.

Taking  all these hints,  what is the most  natural SUSY model one can find?
The next-to-MSSM (NMSSM) stands in the foreground. Historically, it was proposed as a simple solution to
the $\mu-$problem, which  is also a naturalness problem in the MSSM
 (For details, see Ref.~\cite{Ellwanger:2009dp} and references therein.). As a by product,
it has  an advantage in increasing the tree-level Higgs boson mass. Therefore, it
 allows  the lighter stops and weakens the
correlation between $m_Z$ and $m_h$.  This property not only resolves the LEP crisis,
but also is  likely to be the most promising  savior
of  natural SUSY at the LHC (For recent discussions on the NMSSM with a relatively  heavy  Higgs boson,
see  Ref.~\cite{Hall,Arvanitaki,Ellwanger,hgg}.).

In this work, we study the  naturalness via $m_Z$ or $m_h$ in the MSSM and NMSSM using a bottom-up
approach. Given the light stops and gluino with respectively the LHC lower bounds $\sim200$ GeV and
$600$ GeV~\cite{lightstop}, the least degree of fine tuning involving $m_Z$ is due to the gluino,
and roughly $2.5\%$ for both models.
However, taking the hints for $m_h\sim125$  GeV seriously, we show that the NMSSM is more natural.
We study several scenarios which may give a relatively heavy SM-like Higgs boson in this model:

\begin{itemize}

  \item A large $\ld$ and  the doublet-singlet Higgs mixing effect pushing  $m_h$ upward,
which has a heavier Higgs boson and an significant di-photon excess.   However,
  this scenario has a new large fine-tuning due to vacuum stability and
large $\ld-$ renormalization group equation (RGE) running effects.

\item A large $\ld$ but the mixing effect pulling $m_h$ downward, $\ld\sim0.7$ and
  $\tan\beta\sim2$ allows $m_h\sim125$ GeV. An essential difference between
this scenario and the above previous  is the absence of a lighter Higgs.
In addition, it is very difficult to give the significant di-photon excess.

\item A  small $\ld$ and  the mixing effect pushing  $m_h$ upward. We do not have
significant di-photon excess, and the naturalness status becomes worse.

\end{itemize}

In all the above scenarios the neutralino LSP DM candidate
is strongly disfavoured by the XENON100 experiment.  Even if the neutralino LSP DM is fine,
there still exists fine-tuning.
Note that the above analyses are based on the mSUGRA-like model with the mediation scale $M=M_{\rm GUT}$,
but when $M$ is sufficiently low the naturalness can be improved by about one order.

This paper is organized as follows.  In  Section II,  we make  a detailed analysis on the naturalness
implication via $m_Z$ and $m_h$ in the MSSM and NMSSM. We discuss the related phenomenological
consequences  in Section III.
 The Section IV is the discussion and
conclusion. Some necessary and complementary details are given in Appendices A and B.

\section{The Road to the Most  Natural MSSM and NMSSM}

The LHC is testing supersymmetric models. The most predictive model such as
the Constrained MSSM (CMSSM) has been pushed to the multi-TeV region directly by the
SUSY search or indirectly by the Higgs search.
Naturalness is seriously challenged there, nevertheless the NMSSM still can be natural.
In this Section, we will study the origin of fine-tuning in the MSSM and NMSSM
via the bottom-up approach.

\subsection{Light stops and gluinos: The natural soft SUSY spectrum for $m_Z$}

We will first discuss the naturalness implication on the Higgs sector with
successful electroweak (EW) symmetry breaking defined at the weak scale,
and then use RGEs to extrapolate relevant soft parameters to the UV boundary and
examine the naturalness in terms of fundamental soft parameters.
% where the fundamental SUSY breaking soft terms are introduced.

\subsubsection{Naturalness of the electroweak Higgs sector}\label{natural:EW}

The Higgs parameters not only determine the EW-scale $\sim m_Z$ but also have close relation
with the SM-like Higgs boson mass $m_h$. In the MSSM, the tree-level Higgs quartic coupling
is uniquely determined by the $SU(2)_L \times U(1)_Y$ gauge couplings via D-terms, which
implies the tree-level upper bound on the lightest CP-even Higgs boson mass:
$m_h^2\leq m_Z^2\cos^22\beta$. To make up the gap between $m_Z\cos\beta$ and the hinted mass
$m_h\simeq125$ GeV, a substantial radiative correction from the top-stop sector is necessary, i.e.,
 \begin{align}\label{1loop}
\delta_t^2\simeq\f{3m_t^4}{4\pi^2 v^2}\L \log\L\f{m_{\wt t}^2}{m_t^2}\R+\f{X_t^2}{m_{\wt t}^2}\L1-\f{X_t^2}{12m_{\wt t}^2}\R\R,
  \end{align}
  with   $X_t\equiv A_t-\mu\cot\beta$ the stop mixing and $m_{\wt t}$ the geometric mean of
two stop masses.
To sufficiently lift  $m_h$,  a rather heavy stop sector  with $m_{\wt t}\gtrsim 700$ GeV is
 needed, even in the maximal mixing scenario~\cite{Hall}.
In turn we are forced to accept a large $-m_{H_u}^2$ which is implied by  Eq.~(\ref{mHu}).

Therefore, in the MSSM such a heavy Higgs boson is obtained at the price of a
fine-tuned Higgs sector. This can be seen from the following tadpole equation
which determines the weak scale $m_Z$
\begin{align}\label{Mz}
\f{m_Z^2}{2}\simeq&\f{m_{H_d}^2-\tan^2\beta\, m_{H_u}^2}{\tan^2\beta-1}-\mu^2,
  \end{align}
where the parameters are defined at the EW scale. If all the Higgs parameters lied around
$m_Z$ ($m_{H_d}^2$ is an exception whose contribution to the $Z$ boson mass
may be suppressed by large $\tan^2\beta$~\footnote{This manner of the EW symmetry breaking may be
generic in the GMSB with dynamical $\mu/B\mu$  solution~\cite{mHd2}, and
it indeed does not induce the new naturalness problem.}),  then the determination of
 $m_Z$ would be natural.
Otherwise, one can adopt the following quantity to measure
the degree of fine-tuning associated with $m_Z$~\cite{finetuning:def}
\begin{align}\label{finetuning}
\Delta_Z\equiv{\rm max}_iF_i,\quad F_i=\left|\f{\partial\ln m_Z}{\partial\ln p_i}\right|,
   \end{align}
with $p_i$ the fundamental parameters. The above definition can be applied to any quantity which is a consequence
of cancellation. The putative $m_h\sim$125 GeV renders $\Delta_Z<0.1\%$ or even worse~\cite{Hall}.

Turn our attention to the NMSSM, in which the impact of a heavier $m_h$ on the naturalness of $m_Z$
can be abated considerably, since the NMSSM specific effects are capable of lifting $m_h$ without heavy stops.
Nevertheless, $\mu$ as well as other Higgs parameters themselves may hide new fine-tunings,
when we are committing ourself to find a relatively heavy $m_h$. To investigate the actual naturalness status
of the NMSSM (here only the $Z_3-$NMSSM is under consideration),
we start from the Higgs sector
\begin{align}\label{WS}
W=&\ld S H_u\cdot H_d+\f{\kappa}{3}S^3,\cr
 -{\cal L}_{soft}=&m_{H_u}^2|H_u|^2+m_{H_d}^2|H_d|^2+m_{S}^2|S|^2
+\L  \ld A_{\ld}
SH_u\cdot  H_d+\f{{\kappa}}{3}A_{\kappa} S^3+h.c.\R.
  \end{align}
On a proper vacuum we have $v_s=\langle S\rangle\neq0$, therefore the $\mu=\ld v_s$ is generated
dynamically. The non-observation of charginos at the LEP2 gives a bound $\mu>104.5$ GeV.
To demonstrate how new fine-tuning may arise in determining $\mu$, analogously to Eq.~(\ref{Mz})
we trade the order parameter $v_s$ with $\mu$ in the singlet scalar tadpole equation, and get
 %    , which satisfies the equation
 \begin{align}\label{tadpole:s}
\mu\L 2\f{\kappa^2}{\ld^2} \mu^2+\f{\kappa}{\ld} A_\kappa\mu+m_S^2+\ld^2v^2-\ld\kappa v^2 \sin2\beta\R-\f{1}{2}\ld^2 A_\ld v^2 \sin2\beta=0.
  \end{align}
Then if all the NMSSM specific soft parameters lie much above the weak scale, a small
$\mu$ will be a result of fine-tuning.

A remark deserves attention. Tied to $\mu$ and hence the weak scale directly, the chargino
mass is a key to understand the naturalness of SUSY. The absence or the discovery  of a light
chargino $\sim200$ GeV in the future will shed light on it. Interestingly, we will find that
\emph{the naturally heavier $m_h$ and smaller $\mu$ are inherently consistent in the NMSSM}.
Concretely speaking, $\mu$ will be found to automatically fall into the narrow region
100-300 GeV, which ensures the proper doublet-singlet ($H/S$) mixing as well as a non-tachyonic
light stop secctor.

A subtle hidden fine-tuning is associated with vacuum stability. It arises in the
large $\ld$ limit, when the strong RGE effects significantly change the values of soft
parameters, saying $m_S^2$,  during the RGE flowing. We postpone to Section~\ref{lld:cancel}
for the concrete discussions on how does it render the fine-tuning in vacuum stability.

Finally, we have to remind that the $\beta-$angle is dynamically determined through
the third tadpole equation
\begin{align}\label{beta}
\sin2\beta=&\f{2B\mu}{m_{H_u}^2+m_{H_d}^2+2\mu^2+\ld^2v^2},
  \end{align}
which seemingly does not invoke fine-tuning. However, at least in the MSSM this equation can be
transformed to the determination of the CP-odd Higgs boson $A$ mass square:
\begin{align}\label{}
m_A^2=2B\mu/\sin2\beta=m_{H_u}^2+m_{H_d}^2+2\mu^2+\ld^2v^2.
  \end{align}
If $m_A$ were measured to be far below the weak scale, then it would introduce a new
source of fine-tuning comparable to the one of $m_Z$.

\subsubsection{ Feed back to the boundary}\label{UV:natural}

In the previous subsection,  we only outline a natural  Higgs sector at the weak scale.
But its parameters are not fundamental  and
  should be evolved  to the
  mediation scale $M$.  In this work   two  prevalent  mediation scales are considered:
the GUT scale $M_{\rm GUT}$  mediation with $M=M_{\rm GUT}$ in the mSUGRA-like model, and
the low mediation scale with $M=10^5$ GeV in the gauge mediated SUSY breaking (GMSB)
(For a review, see Ref~\cite{GMSB}.). Since we are endeavoring to explore a less fined-tuned
model, some unification assumptions in the mSUGRA model are dropped. Concretely speaking,
the Higgs sector parameters are free, and the stops are treated in a special role just as
in the SUSY framework with the light third family sfermions~\cite{Brust:2011tb}. Howbeit,
the gaugino mass unification is preserved, and similar strategies will be adopted in the GMSB.

SUSY with high mediation scale suffers large fine-tuning, and one of the main object
of this work is to explore how natural it can be. The exploration relies on one fact:
the soft parameters at the low energy SUSY-scale $M_S=1$ TeV are polynomial functions of
the initial values at the GUT scale, with coefficients being functions of the Yukawa and
gauge couplings~\cite{Mambrini:2001wt}. For example, the Higgs doublet soft mass squares
receive the large corrections from gluinos and/or squarks:%高能输入$Y0 = {1.3, 0.001, 0.001, 1.1*4.5, 4.0}/(4 \[Pi])^2$
{\small\begin{align}\label{}
m_{H_u}^2\simeq&-.04  \bar A_t ^2 + .19  \bar A_t   M_{1/2}  - 2.17  M_{1/2}^2 - .11  \bar m_{H_d}^2  +
 .42  {\bar m_{H_u}^2}  - .08 \bar m_S^2  - .33 \bar m_{\wt U^c}^2-.43\bar m_{\wt Q}^2,\\
m_{H_d}^2\simeq&0.02  \bar A_t^2  - .02  \bar A_t   M_{1/2}  + .50  M_{3}^2 + .73  m_{H_d}^2  -
 .14  {\bar m_{H_u}^2}  - .13  \bar m_S^2,
  \end{align}}where  the  parameters with bar are defined at the UV boundary.
 We have set $\ld=0.62,\,\kappa=0.30,\,\tan\beta\simeq2.5,\,m_t\simeq173.4$ GeV at
 $M_S$ (it is in the bulk region studied  numerically later), and these inputs
 will be adopted throughtout this Section.

Now we are able to evaluate the fine-tuning of $m_Z$ with respect of
the various fundamental soft term parameters. The
$Z-$boson mass square is expressed as
\begin{align}\label{natural:mz}
\f{m_Z^2}{2}\simeq&2.68 M_{1/2}^2-.53{\bar m_{H_u}^2}+.27\bar m_{H_d}^2\cr
&+.53\bar m_{\wt Q}^2+.40\bar m_{\wt U^c}^2-.06\bar A_t^2-.22\bar A_tM_{3}+.07\bar m_S^2.
   \end{align}
It is supposed that $\mu\sim100$ GeV does not contribute much to the fine-tuning of $m_Z$, so
 it is not included  here. Then the possible large tuning, in light of Eq.~(\ref{finetuning}),
 are calculated to be
\begin{align}\label{}
& F_{M_{1/2}}\simeq2\times2.68M_{3}^2/m_Z^2,\quad F_{{\bar m_{H_u}^2}}=1\times0.53{\bar m_{H_u}^2}/m_Z^2,\quad
F_{\bar A_t}=2\times 0.06\bar A_t ^2/m_Z^2,  \cr
&F_{\bar m_{\wt U^c}^2}=1\times0.40\bar m_{\wt U^c}^2/m_Z^2,\quad F_{\bar m_{\wt Q}^2}=1\times0.53\bar m_{\wt Q}^2/m_Z^2,\quad F_{\bar m_S^2}=1\times0.06\bar m_{S}^2/m_Z^2,...
   \end{align}
where the dots denote irrelevant terms. Each $F$, the degree of fine-tuning, is
 the \emph{fine-tuning coefficient} (e.g., 0.53 for $F_{{\bar m_{H_u}^2}}$) which encodes the RGE effect,
 times the initial value over $m_Z^2$. Note that we are regarding ${\bar m_{H_u}^2}$ rather than
 $\bar m_{H_u}$ as fundamental. Thus, the corresponding tuning is reduced by half. Otherwise,
 the value should be doubled.

Now we investigate  the naturalness implications
on  various SUSY breaking soft parameters.
\emph{The gluino mass tends to be the dominating source of fine-tuning, as a consequence of the long RGE
running from the GUT-boundary}. The LHC SUSY search places a bound on the gluino mass $M_{\wt g}>600$ GeV
corresponding to the initial value  $M_{1/2}\approx250$ GeV (fixed hereafter), which gives the least
degree of fine tuning:
\begin{align}\label{}
\Delta_Z\geq F_{M_{1/2}}\simeq 40.
   \end{align}
 Throughout this work we will take it as a referred value of the degree of fine-tuning for the natural NMSSM,
 unless otherwise specified. Note that from Eq.~(\ref{beta}) we obtain
 $0.27 {\bar m_{H_u}^2}+0.62{\bar m_{H_d}^2}\sim 2B \mu/\sin2\beta\sim10^5$ GeV$^2$, where
 $B\simeq A_\ld\sim 400$ GeV. So typically we have
 ${\bar m_{H_u}^2}\gtrsim \bar m_{H_d}^2(\sim \bar m_S^2)\sim 10^6$ GeV$^2$, which implies
 that the ${\bar m_{H_u}^2}-$term usually contributes the largest tuning to get the weak scale.

In the light stop framework, it is supposed that stops do not lead to large fine-tunings.
However, the multi-TeVs $\bar A_t$ also results in an enormous fine-tuning and roughly gives the bound
$|\bar A_t |<1.7$~TeV to satisfy $F_{\bar A_t}<F_{M_{1/2}}$. On the other hand, it is likely
that a large $|\bar A_t|$ is necessary to achieve a large mixing in the stop sector,
which is important to enhance $m_h$ with stops as light as possible. To see that clearly,
we consider the following equations:
\begin{align}\label{squark:lld}
 A_t\simeq& 0.19 \bar A_t -0.06\bar A_\ld-1.83 M_{1/2},\\
m_{\wt U^c}^2\simeq&0.61
\bar   m_{\wt U^c}^2 - 0.26\bar   m_{\wt Q}^2-0.04 \bar A_t ^2  + 0.14 \bar A_t    M_{1/2} + 3.04  M_{1/2}^2 -0.20\bar m_{H_u}^2+ 0.03  \bar m_S^2 ,\\
m_{\wt Q}^2\simeq&0.84\bar m_{\wt Q}^2-0.13\bar m_{\wt U^c}^2 -0.02 \bar A_t  ^2 + 0.07  \bar A_t    M_{1/2}+ 4.33  M_{1/2}^2  -0.13\bar m_{H_u}^2+ 0.02 \bar m_S^2.
  \end{align}
As one can see, $-\bar A_t \sim{\cal O}(1)$ TeV helps to achieve a large stop mixing by not only
giving a large $A_t$ but also reducing the stop mass squares.

Comments are in orders. In the first, it can be explicitly seen from Eq.~(\ref{Mz}) that the $m_Z^2$ strong gluino
mass dependence from ${ m_{H_u}^2}$ (as well as the weaker dependence from $m_{H_d}^2$) is exacerbated
as $\tan\beta$ becomes smaller, so a small $\tan\beta\lesssim2$ is not favored for getting the optimal $m_Z$.
Next, the previously discussed fine-tunings involving $m_Z$ originate from the stop and gluino sector,
but are independent on the singlet sector. So the conclusions can be applied to both the MSSM and NMSSM.

Due to the large $\ld$, the soft terms in the NMSSM singlet sector develop strong dependence
on the MSSM terms, e.g., the Higgs soft terms. That may give rise to new fine-tunings.
The singlet soft terms are given by
\begin{align}\label{A:Sld}
m_S^2\simeq&
  % 0.02 A_\ld^2 + 0.26 m^2_0   + 0.05 M_{1/2}^2
- 0.30\bar  m_{H_d}^2 -
   0.21  {\bar m_{H_u}^2}+0.10\bar m_{\wt Q}^2+ 0.25 \bar m_S^2+...,
\\
A_\ld\approx &-0.24 \bar A_t +0.31\bar A_\ld+0.27 M_{1/2},\cr
A_\kappa\approx &0.31 \bar A_t+0.81\bar A_\kappa-1.06\bar A_\ld+0.06 M_{1/2}.\label{A:Sld1}
  \end{align}
Several  remarks about the RGE effects on the low energy soft terms are in orders:
(A) $|m_S|$ develops a strong dependence on $|m_{H_{u,d}}|\sim 1$ TeV, and thus it is typically
close to the TeV scale in the absence of fine-tunings. On the other hand,
in Section~\ref{lld:cancel} we will find that vacuum stability typically requires
$|m_S|<100$ GeV. As a consequence of this tension, $m_S^2$ usually renders a rather large fine-tuning
\begin{align}\label{ms:tuning}
\Delta_S\simeq 0.25 \bar m_S^2/m_S^2 .
  \end{align}
(B) The fine-tuning coefficient of $\bar m_S^2$ is self-decreased by a large $\kappa$.
Varying $\kappa$ from 0.1 to 0.3, it decreases about three times (with $\ld$ fixed). We will use a
simple function to fit this behavior, which is good enough to calculate $\Delta_S$ numerically.
(C) Similarly, the $\bar A_\ld$ and $\bar A_\kappa$ components in $A_\kappa$ and $A_\ld$ are sensitive to
$\ld$ and $\kappa$. Especially, in the limit $\ld\ra0.7$, the $\bar A_\ld$ component increases much
in $A_\kappa $ whereas self-squeezes substantially in $A_\ld$. Therefore, the expected
natural order is $|A_\kappa |>|A_\ld|$, otherwise a new fine-tuning may occur.

Now we turn the attention to the GMSB with low mediation scale. In contrast to the high mediation scale case,
for $M=10^5$ GeV the gluino effect is reduced and the fine-tuning is thus considerably improved.
Concretely, now the $Z$ boson mass square assumes a form
\begin{align}\label{}
m_Z^2\simeq2\L0.08 M_{3}^2+0.17\bar m_{\wt Q}^2+0.15\bar m_{\wt U^c}^2+0.12\bar A_t^2...\R.
   \end{align}
Compared to Eq.~(\ref{natural:mz}), it is clearly seen that the gluino mass dependence has been
reduced one order of magnitude, although the squark influence is still rather significant.
Thus again light stops with masses $\sim600$ GeV, are elements for natural SUSY.
As a comparison, the squark soft mass squares are now given by
\begin{align}\label{}
m_{\wt U^c}^2\simeq&  0.91  \bar m_{\wt U^c}^2-0.08\bar m_{\wt Q}^2  + 0.41  M_3^2-0.06 \bar A_t^2,\\
m_{\wt Q}^2\simeq& -0.04  \bar m_{\wt U^c}^2+0.95\bar m_{\wt Q}^2  + 0.44  M_3^2-0.03 \bar A_t^2,\\
A_t\approx& 0.72 \bar A_t -0.35 M_{3}.
  \end{align}
As for the singlet sector, because of the much shorter RGE running, the $\ld-$RGE effect is reduced considerably
or even ignorable and then not listed.

At the messenger boundary, we have assumed that the stop mass squares are naturally and properly small. Thus,
in principle the fine-tuning can be reduced to a level of 10$\%$ or even better, depending on the gluino
mass ($M_{\wt g}$=750 GeV to get 10$\%$). However, in the minimal GMSB the gluino mass is tied with the
first and second family squark masses. The exclusion of such light squarks means that $M_{\wt g}$ should lie
above the TeV scale. Consequently $\Delta_Z$ is not quite optimistic again. But recalling that generating
acceptable large gaugino masses is a generic problem in the GMSB with dynamical SUSY-breaking~\cite{gaugino}, we conjecture
that the gaugino masses may be suppressed comparing to the squark soft masses. Then again $\Delta_Z$ is directly
related to the LHC gluino search, and $\Delta_Z\sim 20\%$ is still viable from the present data.
In a word, naturalness strongly prefers the GMSB-like models.

To end up this section, we summarize the naturalness implications on the GUT-scale soft
terms for high scale mediation:
\begin{itemize}
  \item The stops and gluino should be as light as possible. Moreover, $\bar A_t$
 enters $m_Z$ and its rough naturalness bound is $|\bar A_t |\lesssim1.7$ TeV.
   \item By virtue of a large $\ld$, one expects that the low energy soft terms show
   $m_S^2$ $\gtrsim {\cal O}(10^5)$ GeV$^2$ and $|A_\ld|<|A_\kappa|$. Otherwise, extra fine-tunings are introduced.
\item In the weak coupling limit the $\ld-$RGE effect is reduced greatly, so the NMSSM does not introduce extra fine-tuning.
\end{itemize}
If the mediation scale is sufficiently low, naturalness of both the $Z$ boson and Higgs boson masses
 can be improved much. In light of the previous analysis, \emph{the inventory of the natural NMSSM in the deformed GMSB is:
the properly suppressed gaugino (at least gluino) masses, light stops, and of course, a sufficiently low messenger scale}. We leave this more
 optimistic scenario for a future work~\cite{new}. In what follows we will focus on the high mediation scale case.

\subsection{Mixing scenario: a natural heavy Higgs in the NMSSM}\label{PQ-INMSSM}

In the previous subsection we have shown that on the naturalness of $m_Z$ alone, the LHC
imposes the common lower fine-tuning bound for both models. Thus the NMSSM does not have
an obvious advantage over the MSSM, and maybe it is even more tuned due to a large $\ld$.
However, it is preferred when taking the Higgs boson mass into account.
In the LEP era, the NMSSM alleviates the little hierarchy problem via two different strategies:
Pushing $m_h$ above 114.4 GeV with a large $\ld$ and a small $\tan\beta$~\cite{Llambda} or
a proper $H/S$ mixing effect~\cite{Dermisek:2007ah}~\footnote{We thank S. Chang informed
us of their earlier study on the mixing effect in Ref.~\cite{Chang:2005ht}.}, or alternatively
allowing $m_h$ below the LEP bound but requiring non-standard Higgs decays to escape from
the LEP search~\cite{HAA}. Now the ATLAS and CMS hints exclude the second scheme, but the
first scheme still works and may be favored viewing from the Higgs di-photon excess.

Within our knowledge, previous literatures on the naturalness of $m_h$ are based on the
LEP bound, and treat the $H/S$ mixing effect mainly using a numerical method~\cite{Dermisek:2007ah,Ellwanger:2011mu}.
Thus intuitions may get lost. As one of the main object of this work, we employ a
detailed analysis of this effect using a semi-analytical method. We start from the
following two by two mass squared matrix
\begin{align}\label{HS:22}
M_{H/S}^2=\L\begin{array}{cc}
               m_Z^2\cos^22\beta+\ld^2v^2\sin^22\beta+\delta_t^2 & \f{1}{2}(4\ld^2v_s^2-M_A^2\sin^22\beta-2\ld\kappa v_s^2\sin2\beta)\f{v}{v_s}\\
                & \f{1}{4}M_A^2\sin^22\beta\f{v^2}{v_s^2}+4\kappa^2v_s^2+\kappa A_\kappa v_s-\f{\ld}{2}\kappa v^2\sin2\beta
             \end{array}\R,
\end{align}
which approximately encodes the $H/S$ mixing effect in light of Appendix~\ref{CPeven:Higgs}.
We denote its eigenstates as $H_{1,2}$ with mass ascending. $(M_{H/S}^2)_{11}$ gives the well-known
NMSSM upper bound on the SM-like Higgs boson mass without taking the mixing effect into consideration.
We are ignoring the correction from the $13-$mixing effect, which decreases $(M_{H/S}^2)_{22}$ by an
amount $(M_{S}^2)_{13}^2/M_{11}^2$. Since $(M_{S}^2)_{13}\propto \cos2\beta$. The decrease may be
appreciable for a large $\tan\beta$, but it will not affect our main conclusion. So we leave this
subtle effect to numerical studies. In addition, the top-stop loop correction gives
$\delta_t\simeq {\rm 50\, GeV}-{\rm 80\, GeV}$ for $m_{\wt t}\simeq 300$ GeV and a moderate mixing effect.
Throughout our analysis, such a referred value will be used as the premise of defining natural SUSY.

\subsubsection{A general analysis on the $H/S$ mixing effect}

As a general discussion on the mixing effect and its implication, we start from a general
matrix structure rather than confining to the NMSSM. So our discussion can be applied to many
new physics models showing a $H/S$ mixing, e.g., the SM Higgs sector extended with an extra
scalar singlet in the context of dark matter models~\cite{Gao:2011ka,Kim:2008pp}. For an arbitrary
2$\times2$ symmetric real matrix $M^2$, \emph{the necessary and sufficient condition that
the mixing effect pushes the larger diagonal element $M_{11}^2$ to an even larger eigenvalue
(namely the SM-like Higgs boson mass)} is
 \begin{align}\label{HS:conditions}
M_{11}^2\geq M_{22}^2, M_{12}^2, \quad M_{11}^2M_{22}^2\geq M_{12}^4.
  \end{align}
 The second equation is the minimum condition or vacuum stability condition. For our purpose,
 a natural $M_{11}$ should be around 120 GeV. The two eigenvalues of $M^2$ are written as
 \begin{align}\label{}
m_{1,2}^2=\f{1}{2}\L \Delta_+\pm 2M_{12}^2\left[\L\f{ \Delta_-}{2M_{12}^2}\R^2+1\right]^{1/2}\R,
  \end{align}
where $\Delta_\pm= M_{11}^2\pm M_{22}^2$. The mixing effect pushes upward the ``mass"
(in the flavor basis) of the doublet $M_{11}^2$ to the eigenvalue $m_{1}^2$, whereas pulls
downward the singlet ``mass"  to $m_{2}^2$. We will use the pushing or pulling effect to
describe these two faces of the mixing effect for short.

The implications of the mixing effect are two-folds: One is the mixing angle between states,
and the other is the pushing or pulling effect on ``Higgs boson masses".
The unitary rotation diagonalizing the mass square matrix is given by
 \begin{align}\label{HS:mixing}
U=\L\begin{array}{cc}
               \sin\theta &\cos\theta\\
               -\cos\theta & \sin\theta
             \end{array}\R,\quad \tan\theta=\f{\Delta_-}{2M_{12}^2}+\left[ 1+\L\f{\Delta_-}{2M_{12}^2}\R^2\right]^{1/2}.
  \end{align}
Thus the mixing angle is determined solely by the ratio $\Delta_-/2M_{12}^2$, which can be large
even for a very small off diagonal element $M_{12}^2\ll M_{11}^2$. Hence we conclude: Although
the mixing angle is of importance in the Higgs collider phenomenology, it can not properly measure the amount
of pushing upward or pulling downward $m_h$.

To estimate the size of the enhancement effect, we go to some special but well-motivated limits:
\begin{description}
\item[Decoupling limit]  When the off diagonal element is so small that $2M_{12}^2\ll |\Delta_-|$, it
is appropriate to define the decoupling limit where both the mixing angle and enhancement is
vanishing small. In this limit the doublet mass eigenvalue is given by
  \begin{align}\label{}
m_{1}^2\approx  M_{11}^2\L 1+\f{1}{4}\times\f{2M_{12}^2}{M_{11}^2}\f{2M_{12}^2}{\Delta_-}\R.
  \end{align}
Since $2M_{12}^2/M_{11}^2<{2M_{12}^2}/{\Delta_-}<0.3$ by definition, the enhancement is indeed
ignorably small in the decoupling limit: $m_{11}-M_{11}<1\% M_{11}$.
   \item[Marginal push]  Considering $M_{11}^2\gg M_{22}^2$, one can imagine that the mixing effect
   should be marginal for a properly large $M_{12}^2$. In this case the heavy Higgs boson mass
   square is approximated to be
  \begin{align}\label{}
m_{1}^2\approx M_{11}^2\left[1+\f{M_{12}^4}{M_{11}^4}\L1+\f{ M_{22}^2}{M_{11}^2}\R\right],
  \end{align}
and the mixing angle is $\tan\theta\sim M_{11}^2/M_{12}^2>|M_{11}/M_{22}|$ by Eq.~(\ref{HS:conditions}).
For $M_{11}=120$ GeV, the marginal mixing effect is able to give the necessary push, e.g.,
$M_{11}^2=3M_{22}^2$ and $M_{12}^4=M_{11}^4/8$. While $M_{11}=115$ GeV does not work.
 \item[Significant push] The large enhancement effect will happen when $M_{11}^2\gtrsim  M_{22}^2$ and
 $M_{12}^2\gg |\Delta_-|$. Then we have
  \begin{align}\label{}
m_{1}^2\approx  \f{1}{2}\L M_{11}^2+M_{22}^2\R+M_{12}^2+\f{\L M_{11}^2-M_{22}^2\R^2}{8M_{12}^2}.
  \end{align}
  In principle, the mixing effect could be able to make the heavier Higgs approaching the maximum value
  $\simeq\sqrt{2}M_{11}$, together with a light Higgs boson. The mixing angle in this case is large.
\end{description}

\begin{figure}[htb]
\begin{center}
\includegraphics[width=3.2in]{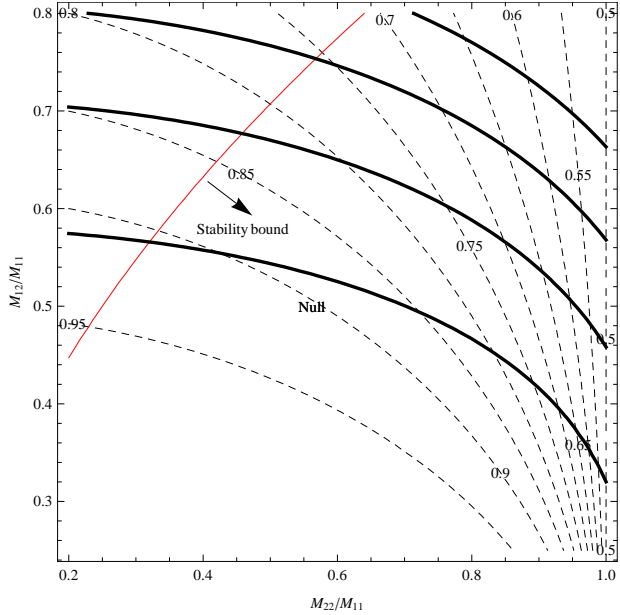}
\end{center}
\caption{\label{HS} The contour plot of the enhancement factor $m_{1}/M_{11}$ (thick black lines,  $m_{1}/M_{11}=1.05,\,1.10,\,1.15,\,1.20$ respectively from bottom up) and $\sin^2\theta$  the  doublet  component of $h$  (dashed lines) in the $M_{22}/M_{11}-M_{12}/M_{11}$ plane.}
\end{figure}

More generically, the numerical plot in Fig.~\ref{HS} gives some insights into the features of
$H/S$ mixing, confronting with the collider bound:
\begin{itemize}
  \item  For a fixed pushing strength (saying 1.05), the doublet fraction varies in the region
  $\sim$0.50-0.90. For the fraction $\simeq0.5$, the diagonal elements are quite degenerate,
  which implies that the two eigenvalues are roughly equal. In addition, such a large mixing
  renders the definition of a SM-like Higgs boson not very clear. And generically the $h\ra 2\gamma$
  rate is reduced and then not favorable. If the fraction is 0.9, the heavier Higgs field $h=H_2$
  pairs with a lighter singlet-like Higgs boson $H_1$, whose mass is typically below the naive LEP upper bound.
But it is likely that at some intermediate value of the doublet fraction, aided by a large enough $M_{22}$,
the lighter Higgs $H_1$ is allowed by the LEP.
  \item We can also consider the pulling scenario in which the large $\ld$ and small $\tan\beta$ have
  already lifted $m_h$ up to $\sim125$ GeV. Thus, we have to preclude the pulling effect from decreasing
  $m_h$ too much. This scenario is phenomenologically distinctive owing to the absence of a new light Higgs boson
  other than $h=H_1$.
\end{itemize}
In the following sections, we will  explore the natural Higgs scenarios along these lines.

%%%%fig.2%%%%%%%%%%%%%%%%
%%%%%%%%%%%%%%%%%%%%%%%%%%%%%%%%%%%%%%%%%%%%%%%%%%%%%%%%%%%%%%%%%%%

\subsubsection{Large $\ld$: a no-go for the decoupling limit}

The NMSSM with a large $\ld\sim0.7$ and a small $\tan\beta$ can lift $m_h$ to 125 GeV
even without a mixing effect and large loop-correction. However, it does not guarantee
a natural NMSSM. As argued in the previous subsection, $\tan\beta\sim1.7$ mildly exacerbates the
fine-tuning involving $m_Z$. On top of that, a new fine-tuning may emerge when we take
the RGE effects into account.

A large $\ld$ tends to pull the SM-like Higgs boson mass rather than push. To show this, we
rewrite $M_{12}^2$ which is given in  Eq.~(\ref{HS:22}) as
  \begin{align}\label{M12}
M_{12}^2=2\ld\mu v\left[1-\L\f{A_\ld}{2\mu}+\f{\kappa}{\ld}\R\sin2\beta\right].
  \end{align}
If there is no substantial cancellation, the first term $2\ld\mu v$ alone shows that
$M_{22}^2>M_{11}^2$ is needed to avoid tachyon even if $\mu$ tracks the lower bound
$\sim$100 GeV. Consequently Eq.~(\ref{HS:conditions}) is violated and the mixing effect
pulls down the SM-like Higgs boson mass.

To minimize the pulling effect, we resort to the decoupling limit. It requires a quite
heavy singlet sector, which will be found to be impossible. To see it, we consider
the explicit expression for $M_{22}^2$:
   \begin{align}\label{M332}
M_{22}^2={\ld^2 v^2}\f{A_\ld}{2\mu}\sin2\beta+4\f{\kappa^2}{\ld^2}\mu^2+\f{\kappa}{\ld}A_\kappa \mu.
  \end{align}
At first, the large $\ld$ and putative small $\mu$ negate the possibility of % 用that的话后面是从句
the second term to exceed the weak scale much. Next, if we count on $A_\ld \sin2\beta/2\mu\gg1$,
from Eq.~(\ref{M12}) and Eq.~(\ref{M332}) we know that it will render the CP-even Higgs boson mass matrix
tachyonic. Finally, the CP-even and CP-odd singlet mass squares receive opposite contributions from the
$\kappa A_\kappa$ term (see Appendix~A), so the $\kappa A_\kappa$ dominance in $M_{22}^2$ is also excluded.
Eventually we affirm the no-go in the natural NMSSM with a large $\ld$ and whatever value of $\tan\beta$:
\emph{A heavy and decoupling singlet sector is impossible and large $\kappa A_\kappa$ is excluded as well}.
In other words, a proper cancellation is inevitable, and we will discuss the naturalness of this cancellation
in the following.

\subsubsection{Large $\ld$: push versus  pull}\label{lld:cancel}

Previously, it is found that although a large $\ld$ effectively enhances the Higgs boson
mass, cancellation is necessary. So we have to contemplate whether or not the cancellation is
tolerated by naturalness. In light of Eq.~(\ref{M12}) we define
\begin{align}\label{natural:cancel}
C_A\equiv\L\f{A_\ld}{2\mu}+\f{\kappa}{\ld}\R\sin2\beta,
\end{align}
which should be quite close to 1 to sufficiently reduce $M_{12}^2$ through cancellation.
To maintain perturbativity up to the GUT-scale, $\ld>0.6$ means that $\kappa$ should be
moderately smaller than $\ld$. Then $A_\ld/2\mu$ plays the primary role for cancellation
(But $\kappa$ also plays an important role if this ratio is relatively small). So we have
\begin{align}\label{Ald}
A_{\ld}\simeq \f{2\mu}{\sin2\beta}.%\ra  C_A\,\mu\tan\beta \,\quad ({\rm for}\,\tan\beta>3).
  \end{align}
As one can see, the absence of light charginos requires a large $A_\ld$, especially when
$\tan\beta$ is relatively large.

\begin{figure}[htb]
\begin{center}
\includegraphics[width=3.2in]{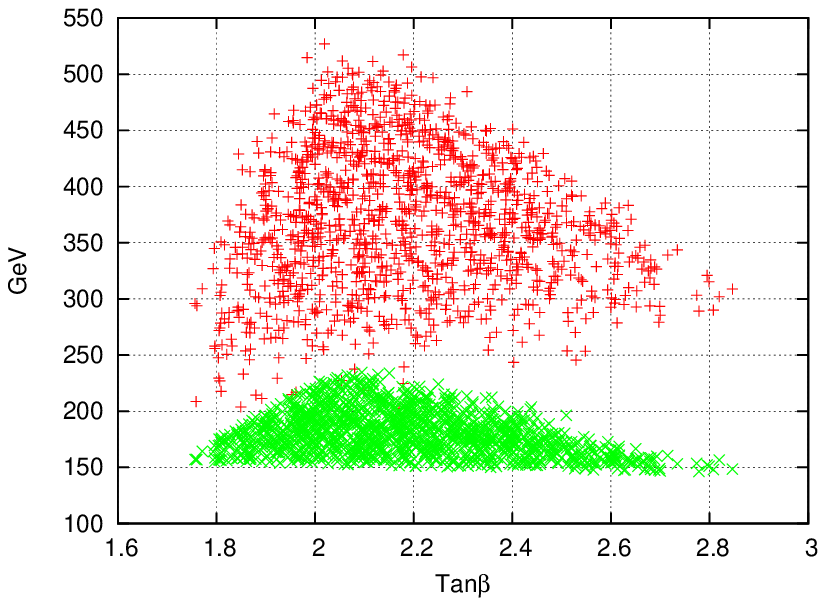}
\includegraphics[width=3.2in]{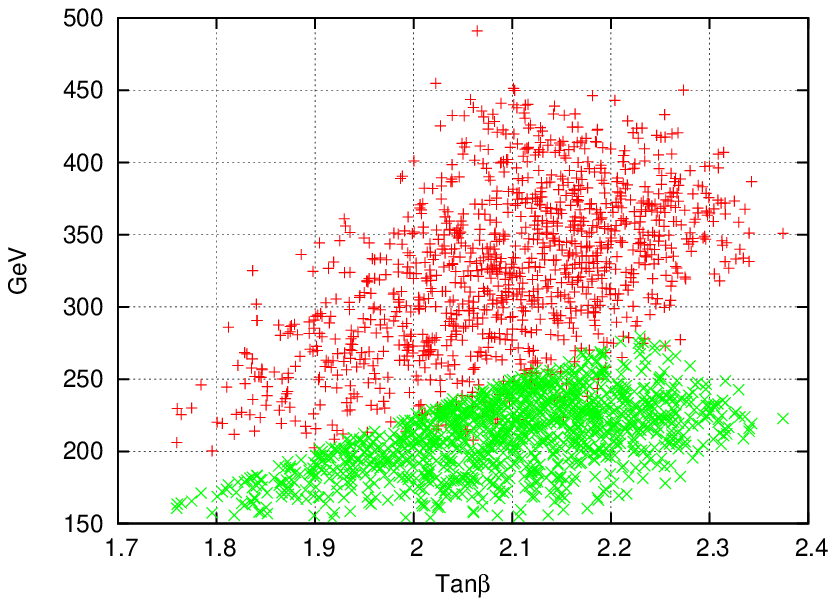}
\includegraphics[width=3.2in]{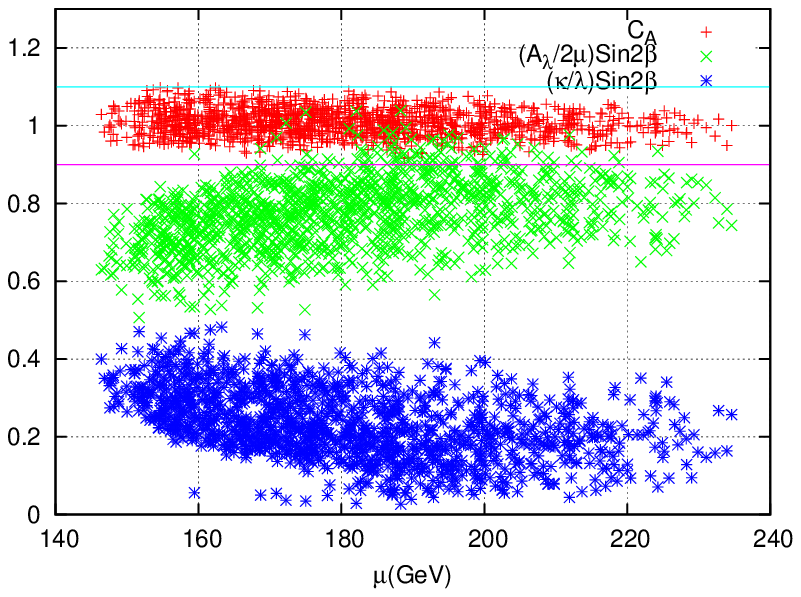}
\includegraphics[width=3.2in]{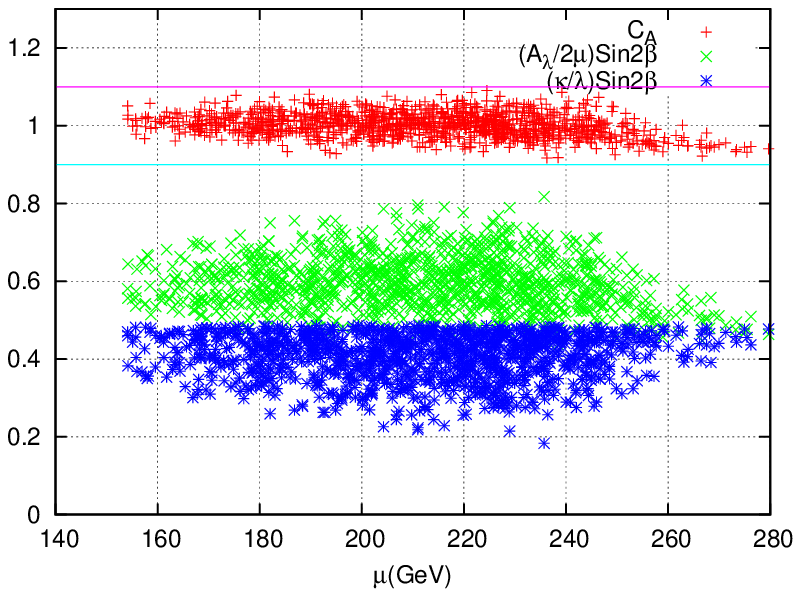}
\end{center}
\caption{\label{Largeld} Predicative natural NMSSM parameter space. Top-left:
$\tan\beta-A_\ld$ (red) and $\tan\beta-\mu$ (green) plots in the pushing scenario. Bottom-left:
Plots of the three ratios  as  functions of  $\mu$ in the pushing scenario. Top-right and bottom-right:
Corresponding plots in the pull scenario.  $\ld=0.60$ is fixed, while the other parameters
and physical constraints are
given in  Section~\ref{numerical}.}
\end{figure}

The parameter space is quite predictive once naturalness and Higgs boson mass conditions are imposed.
First of all, $\tan\beta\sim2$ and $\ld\sim0.7$ are necessary for a relatively heavy SM-like Higgs boson.
Next, we require $\mu\sim200$ GeV for the sake of significant mixing effects and naturalness. Then
$A_\ld$ falls into a narrow region around 400 GeV due to Eq.~(\ref{Ald}). Those observations are confirmed
by Fig.~\ref{Largeld}. Finally, $\kappa$ should be properly large so as to suppress or forbid
the invisible decay of Higgs into a light neutralino pair.

The aforementioned cancellation itself does not mean fine-tuning, since $C_A$ can be determined dynamically
by the singlet tadpole equation Eq.~(\ref{tadpole:s}). From this equation we get
\begin{align}\label{tuning:CAN}
C_A-1=\f{m_S^2+2(\kappa/\ld)^2\mu^2+(\kappa/\ld)A_\kappa \mu}{\ld^2 v^2},
  \end{align}
where $\ld v\simeq 100$ GeV. Therefore, it seems that the right-handed side can be naturally at
$\sim0.1$ due to a sufficiently small $|m_S|$, with the other terms suppressed by small enough
$\kappa$ and $A_\kappa$. But this does not address the naturalness problem correctly if the RGE effects
are taken into account. In fact, $|m_S|\lesssim30$ GeV leads to fine-tuning worse than $1\%$,
which is manifested in Eq.~(\ref{A:Sld}). So the really natural $|m_S|$ should be properly large, e.g.,
$|m_S|\sim10^2$ GeV. As a result, Eq.~(\ref{tuning:CAN}) implies a larger $2(\kappa/\ld)^2\mu^2$,
which forces $\kappa\sim\ld$ (So the PQ-limit is disfavored). On the other hand, a larger $\kappa$ further
improves naturalness by squeezing the fine-tuning coefficient of $\bar m_S^2$. These points will be
confirmed numerically.

Another source of fine-tuning may creep in. We have shown that a large $|A_\ld|$ and a small
$|A_\kappa|$ is favored to avoid tachyonic Higgs states. Thus, we expect the order $|A_\ld|>|A_{\kappa}|$
for vacuum stability. However, in Section~\ref{UV:natural} it was found that the natural order should be
reversed in the large $\ld$ scenario. Consequently it gives rise to a source of tuning. Fortunately,
even taking $|\bar A_t|<1.7$ TeV into account, naturalness merely places a loosely lower bound on $|A_\kappa|$.
For example, setting $A_\ld=400 \,(600)$ GeV and $\bar A_t =-1.15$ TeV, then $\Delta_{A_\kappa}=40$ means
$|A_\kappa|<30 \,(70)$ GeV, which is always satisfied in our numerical study.

The mixing effect on $m_h$ depends on the singlet mass square $M_{22}^2$, whose different values
lead to different scenarios of lifting the Higgs boson mass. We rewrite $M_{22}^2$
using Eq.~(\ref{tuning:CAN})
  \begin{align}\label{M332}
M_{22}^2\simeq \L C_A-\f{\kappa}{\ld}\sin2\beta\R{\ld^2 v^2}+4\f{\kappa^2}{\ld^2}\mu^2+\f{\kappa}{\ld}A_\kappa \mu.
%M_{22}^2\simeq \L C_A-\f{\kappa}{\ld}\sin2\beta\R{\ld^2 v^2}+4\f{\kappa^2}{\ld^2}\mu^2+\f{\kappa}{\ld}A_\kappa \mu.
  \end{align}
There may be a further small correction due to the 13-mixing effect which reduces $M_{22}^2$.
The first term is positive and $\lesssim M_{11}^2$. The other two terms also play some roles,
since $\kappa$ can not be very small. Then, two different cases arise:
\begin{itemize}
   \item The pushing region is realized when $h=H_2$ and the lighter $H_1$ is singlet-like.
      The LEP bound on $m_{H_1}$ means $M_{22}^2$ should be properly large, which can be seen in
     the top-left plot of Fig~\ref{Largeld}. In the small $A_\ld$ region where the first term of
     $M_{22}^2$ is small, a large $\kappa$ is required. From Fig.~\ref{Largeld} we see that $\mu$ is
     \emph{automatically} bounded by $\mu<240$ GeV (The following pulling scenario has a similar property).
     There are two reasons to understand the smallness of $\mu$, both related to the heavy Higgs:
     (1) A small $\mu$ is good for sufficient mixing; (2) We need light stops, a large mixing $-A_t$ as well as
     a smaller $\tan\beta$, and then $X_t$ would be so large that the color symmetry breaks if $\mu\gg m_Z$.
 \item  $\kappa/\ld$ is relatively large ($\sim0.4$, see Fig.~\ref{Largeld}) and the negative third term
 does not cancel the first two terms. Then $M_{22}^2>M_{11}^2$ and we get the pulling scenario. There, asides
 from the relatively heavy SM-like Higgs boson $h=H_1$, exists a mildly heavier singlet-like Higgs boson $H_2$.
 We do not need to worry about the LEP bound, but should keep $M_{12}^2$ sufficiently small to
 minimize the pulling effect. This can be seen by comparing the distribution of $C_A$ in Fig.~\ref{Largeld},
 where the pulling scenario has much more points crowding around $C_A=1$ than the pushing scenario.
 Note that the pulling scenario usually suffers more fine-tuning with a reason traced back to Eq.~(\ref{tuning:CAN}),
 where a rather large $2(\kappa/\ld)^2\mu^2$ has to be canceled by a larger $| m_S^2|$ and a larger ${m_{H_u}^2}$ thereof.
\end{itemize}

%%%fig.2%%%%%%%%%%%%%%%%
%%%%%%%%%%%%%%%%%%%%%%%%%%%%%%%%%%%%%%%%%%%%%%%%%%%%%%%%%%%%%%%%%%%

\subsubsection{The small $\ld$ scenario}\label{h:smallld}

As mentioned in subsection~\ref{PQ-INMSSM}, a mixing effect only may be able to push the SM-like Higss
boson mass close to 125 GeV. This happens in a distinctive parameter space where $\ld$ is moderately
small and then the tree-level Higgs boson mass reduces to the MSSM case, so a large $\tan\beta$ is required.

To get an overall impression on the feature of the parameter space of this scenario, we again follow
Eq.~(\ref{HS:conditions}). First, $M_{22}^2$ is approximated to be
   \begin{align}\label{}
M_{22}^2\approx&4(\kappa/\ld)^2\mu^2+(\kappa/\ld)A_\kappa\mu.
  \end{align}
In $M_{22}^2$ a term proportional to $\L{\ld}{v}/{2\mu}\R M_{12}^2$ has been ignored, on the ground of
$M_{12}^2<M_{11}^2$ and a small $\ld$. $M_{22}$ should be around 100 GeV for a significant push. Recalling
that a large $\kappa A_\kappa$ is inconsistent with vacuum stability, the unique option for $M_{22}$ is
 \begin{align}\label{natural:2}
\kappa/\ld\sim 0.5\L\f{100\rm\,GeV}{\mu}\R. %\quad A_\kappa\lesssim\L \f{100\rm\,GeV}{\mu}\R\L\f{\ld}{\kappa}\R{100\rm\,GeV}
  \end{align}
This constraint on $\kappa$ has important implication on the singlino mass and then the LSP phenomenology.
Now the non-diagonal element $M_{12}^2=2\ld\mu v(1-C_A)$ can be readily suppressed due to a small $\ld$,
even without turning to a small $1-C_A$ via cancellation. In addition, again, like the pushing scenario with a large
$\ld$, here a small $\mu$ is required to control the sizes of $M_{12}^2$ and $M_{22}^2$.

Although the small $\ld$ scenario requires rather large cancellations between the Higgs parameters to satisfy
the tadpole equations which contain $\ld^2 v^2$, they are not new sources of fine-tuning. As mentioned in Section~\ref{natural:EW},
such cancellations are dynamically implemented and do not determine any particle mass scale. Actually,
in the small $\ld$ case the fine-tuning associated with $m_S^2$ disappears, since the strong $\ld-$RGE effects on $m_S^2$
are removed.

\section{The numerical studies in the natural NMSSM}

The natural NMSSM provides a quite attractive framework from the point view of discovery potential, because of the
lightness of neutralinos, the third family squarks, gluino, as well as Higgs boson. In this Section we present
numerical studies with the aid of the NMSSMTools 3.2.0~\cite{NMSSMTools}, not only affirming the points developed previously
but also providing some important phenomenological consequences.

\subsection{The Higgs boson decay: naturalness and 2$\gamma$ excess}\label{numerical}

First, we study the large $\ld$ scenario. Before the numerical study, we present the setup for the parameter space
in the natural NMSSM. We fix the stop sector at $M_S$ following the criterion of minimizing the fine-tuning from $m_Z$:
 \begin{align}\label{inputs}
m_{\wt Q}=350{\rm\,GeV},\quad m_{\wt U^c}=300{\rm\,GeV},\quad A_t=-500{\rm\,GeV},
  \end{align}
which leads to the lighter stop mass $m_{\wt t_1}\simeq200$ GeV. In terms of Ref.~\cite{lightstop}, the current LHC data
still allow such a light stop, and we will take this lower bound on the stop sector hereafter. Of course, one
can choose other configurations, but our choice is likely to approach the least fine-tuned stop sector with a realistic spectrum.
Other sfermions are assumed to be irrelevant. Although this assumption does not hold water for extremely heavy sfermions which
significantly modify RGEs, we suppose that such effects are sub-leading. To illustrate the sensitive dependence of the degree of
fine-tuning on $\ld$, two cases with $\ld=0.60$ and $\ld=0.62$ will be studied for comparison, whereas $\kappa$, $\tan\beta$,
$\mu,\,A_\kappa$ and $A_\ld$ vary freely. We keep the points satisfying all constraints given in the NMSSMTools except the WMAP
and XENON100 bounds which will be studied specifically. In addition, we require the SM-like Higgs boson mass $m_h>120$ GeV and
its doublet fraction should be larger than $0.8$.

%%%fig.2%%%%%%%%%%%%%%%%
 \begin{figure}[htb]
\begin{center}
\includegraphics[width=3.2in]{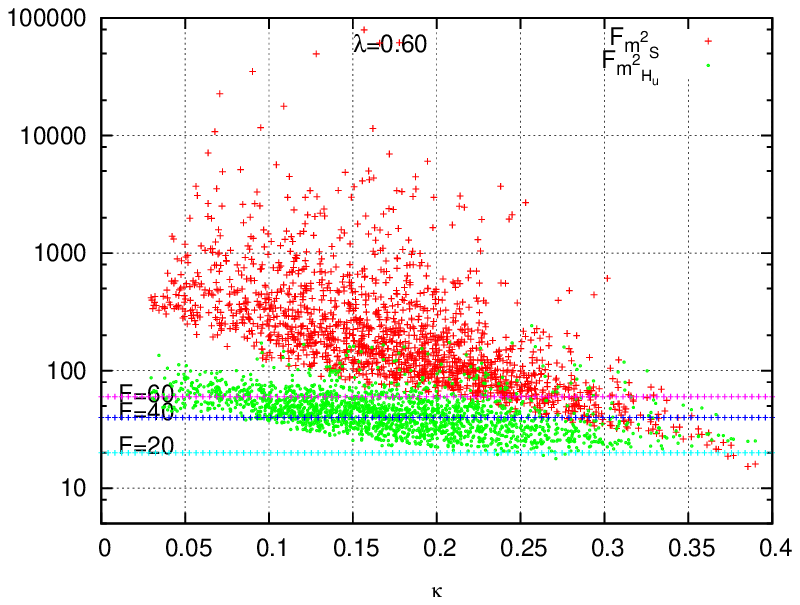}
\includegraphics[width=3.2in]{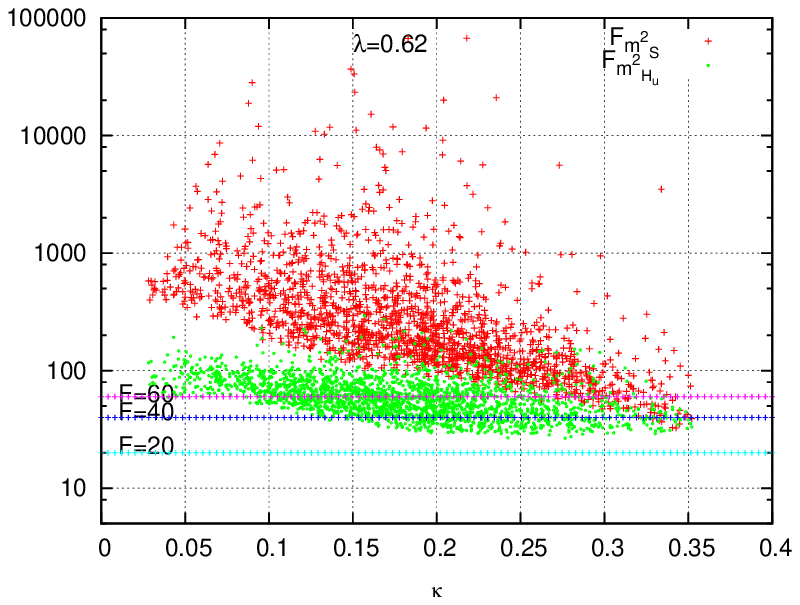}
\includegraphics[width=3.2in]{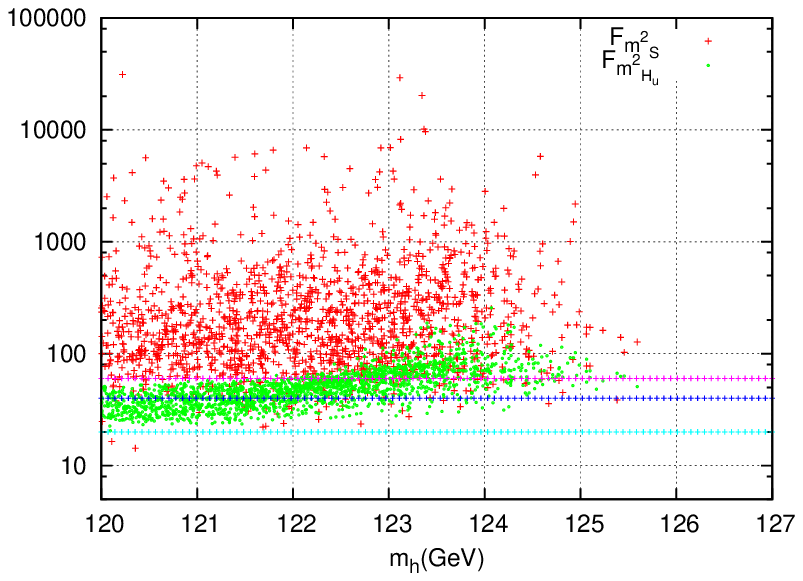}
\includegraphics[width=3.2in]{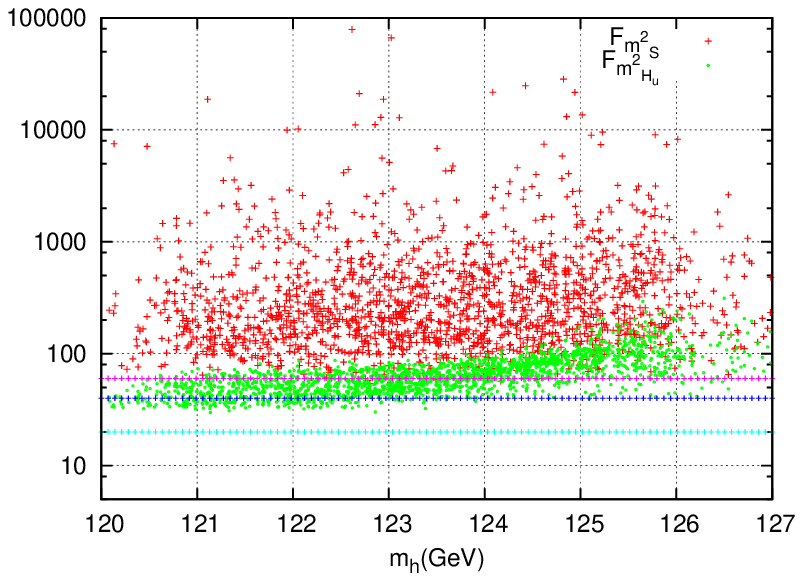}
\end{center}
\caption{\label{tuning} The degree of fine-tuning    $\Delta_{S}$ and   $F_{\bar m_{H_u}^2}$   as functions of  $\kappa$ (top) and $m_h$ (bottom). Left panels: $\ld=0.60$; Right panels: $\ld=0.62$.  Naturalness favors  larger $\kappa$. A heavy Higgs tends to be at the price of larger fine-tuning.}
\end{figure}
%%%%%%%%%%%%%%%%%%%%%%%%%%%%%%%%%%%%%%%%%%%%%%%%%%%%%%%%%%%%%%%%%%%

To calculate the degree of fine-tuning, we record the crucial soft parameters at the GUT-scale boundary, which are obtained
by RGE evolving the parameters from $M_S$ to $M_{GUT}$. The calculation is based on some observations and reasonable approximations:
(1) $F_{{\bar m_{H_u}^2}}$ is the largest degree of fine-tuning involving $m_Z$ (In Eq.~(\ref{natural:mz}) $\bar m_{H_u}^2$ plays
the major role to cancel $M_{1/2}^2$), if it exceeds $F_{M_{1/2}}$. We have the approximation
 \begin{align}\label{}
F_{{\bar m_{H_u}^2}}\approx\f{0.42\times\tan^2\beta+0.15}{1-\tan^2\beta}\f{{\bar m_{H_u}^2}}{m_Z^2},
  \end{align}
which varies only slightly for $\ld=0.62$ and $\ld=0.60$. (2) As discussed in Section~\ref{UV:natural},
$\Delta_S$ defined in Eq.~(\ref{ms:tuning}) is much more complicated than $F_{{\bar m_{H_u}^2}}$, and we use a
function of $\kappa$ to fit the $\kappa$-dependent fine-tuning coefficient. The detailed fitting is shown in
Appendix~\ref{kappa:dep}. With them we can calculate $F$ for each point and plot them in Fig.~\ref{tuning},
from which we obtain:
\begin{itemize}
  \item A large $\ld$ can lift the Higgs boson mass, but count against naturalness.
  In spite of a very small change in the numerical value of $\ld$, the available maximum $m_h$ shifts a few of GeVs.
  However, as $\ld$ increases, the strong $\ld-$RGE effects become more significant and thus $\Delta_S$ is larger.
  If we slightly increase $\ld$ to a value 0.65, $m_h$ can reach $\sim$130 GeV, but $\Delta_S$ typically lies above 100.
  Hence $\ld$ being properly large instead of as large as possible, is favored by naturalness.
  \item The new fine-tuning $\Delta_{S}$ tends to dominate over $F_{\bar m_{H_u}^2}$, except in the large $\kappa$
  region where the $\bar m_S^2$ self-reducing is significant. Nevertheless, a large $\kappa$ strongly prefers the pulling region
  rather than the pushing region as explicitly  explained in  Section~\ref{lld:cancel}. It also interprets the increasing of $F$ with $m_h$.
  \item We are not going to present the fine-tuning in the pulling scenario in this work. It typically gives
  fine-tuning dominated by $F_{\bar m_{H_u}^2}\gtrsim 60$ which is worse than the pushing scenario. But $\Delta_{S}$ is
  never a problem as explained in the above. In this scenario, our results of the minimal fine-tuning $\gtrsim60$ is in
  agreement with the one found in Ref.~\cite{lightstop} which uses the leading logarithm approximation.
\end{itemize}
  In short, we clarify the actual naturalness problem associated with the relatively heavy Higgs boson within the NMSSM
  (with a high mediation scale): \emph{We need a large $\ld$ and a relatively small $\kappa$ to push the Higgs mass,
  but they incur a large fine-tuning in stabilizing the vacuum}. Although $|A_\ld|<|A_\kappa|$ may also destabilize the vacuum,
  we have shown previously that it can be avoided at the price of a small fine-tuning.
 Our results generalize the conclusion of Ref.~\cite{Barbieri:2007tu}, where such kind of new fine-tuning is presented
 only in the PQ-limit.

%%%fig.2%%%%%%%%%%%%%%%%
 \begin{figure}[htb]
\begin{center}
\includegraphics[width=3.2in]{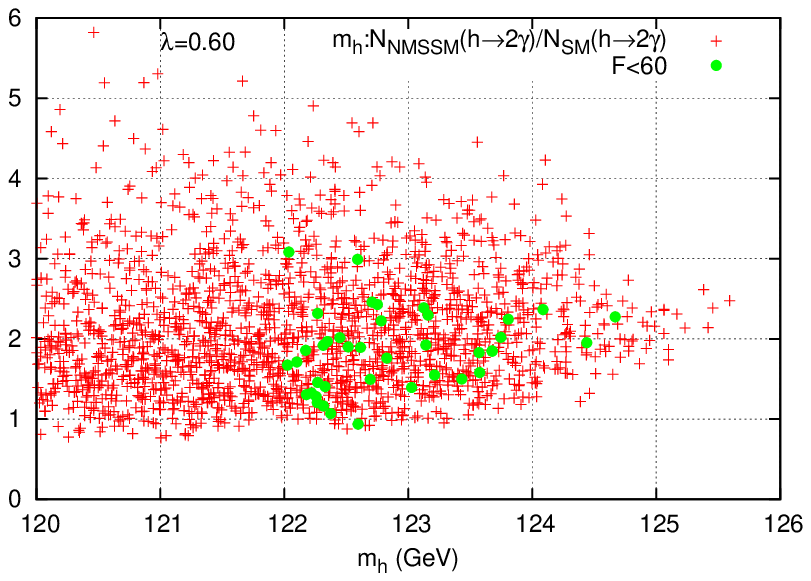}
\includegraphics[width=3.2in]{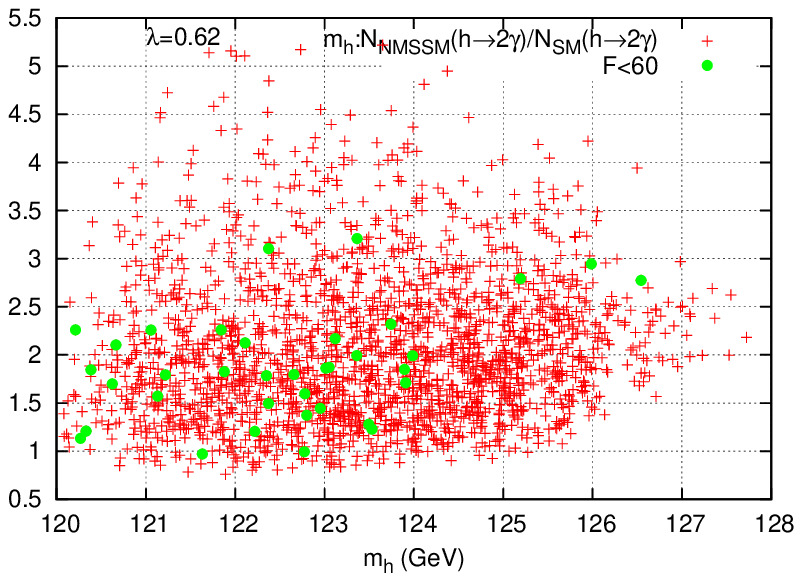}
\includegraphics[width=3.2in]{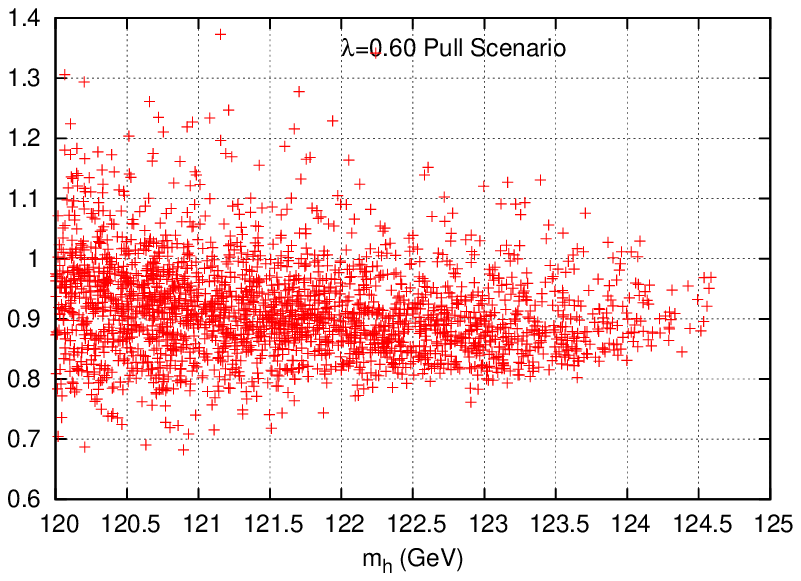}
\includegraphics[width=3.2in]{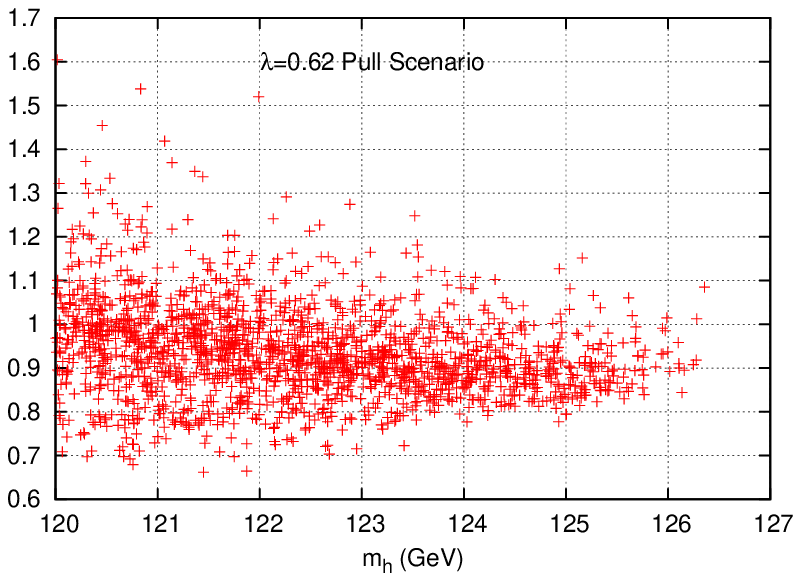}
\end{center}
\caption{\label{h2gamma} The ratio of $h\ra 2\gamma$ between the NMSSM and SM predictions.  Top figures: Pushing
scenarios with  $\ld=0.60$ (left)  and   $\ld=0.62$ (right).
 Bottom figures: Pulling scenarios with $m_{\wt q}=350$ GeV and $A_t=-580$ GeV.}
\end{figure}
%%%%%%%%%%%%%%%%%%%%%%%%%%%%%%%%%%%%%%%%%%%%%%%%%%%%%%%%%%%%%%%%%%%

Now we turn to the Higgs boson collider phenomenology, focusing on the $h \to \gamma \gamma$ signal.
At the ATLAS and CMS experiments, di-photon is the main excess for the hints of the SM-like Higgs boson with mass
in the range 123-127 GeV. Interestingly, to some extent in the natural NMSSM the excess can be regarded as a
prediction from the pushing scenario. To effectively push the Higgs boson mass, the singlet usually
snatches a portion of component from $H_d^0$. Furthermore, the required small $\tan\beta$ may reduce
the bottom Yukawa coupling compared to the large $\tan\beta$ case. As a result, the $h\ra2\gamma$ signal
is likely to be enhanced~\cite{dipho} mainly by reducing the branching ratio BR$(h\ra b\bar b)$. By contrast,
in the MSSM to have such a heavy Higgs boson is already a rather tough task, and the di-photon excess is even more
difficult to get~\cite{Carena:2011aa}. Numerically, we adopt the method used in
Ref.~\cite{hgg} to calculate the ratio of di-photon events in the NMSSM to the corresponding value in the SM:
\begin{align}\label{}
R^{h}(\gamma \gamma)=\frac{\Gamma(gg \to h) \times {\rm BR}(h \to \gamma \gamma)}{\Gamma(gg \to h_{SM})
 \times  {\rm BR}(h_{SM} \to \gamma \gamma)}.
\end{align}
From the NMSSMtools we can extract the decay widths ratio ${\rm BR}(h \to \gamma \gamma)/{\rm BR}(h_{SM} \to \gamma \gamma)$,
and the Higgs production rates ratio $\Gamma(gg \to h)/\Gamma(gg \to h_{SM})$, which is approximated by the square
of the reduced couplings ratio $C^{h}_{gg}=g_{hgg}/g_{h_{SM}gg}$. We plot $R^{h}(\gamma \gamma)$ in Fig.~\ref{h2gamma}, from which
some observations are made:
\begin{itemize}
  \item The pushing scenario shows significant di-photon excess, especially the ATLAS and CMS data can be well fitted
  in the $\ld=0.62$ case. From Fig.~\ref{h2gamma} it is reasonable to conclude: Naturalness prefers the Higgs boson with a
  smaller mass $m_h\sim123$ GeV and $R^h(\gamma\gamma)<2$, which favors the CMS results. In natural SUSY the light stop
  sector affects the di-photon rate from two aspects. On the one hand, it increases the Higgs production cross section
  via gluon fusion to a mount~\cite{Arvanitaki}
\begin{align}\label{}
\f{1}{2}  \L \f{m_t^2}{m_{\wt t_1}^2} +\f{m_t^2}{m_{\wt t_1}^2}-\f{X_t^2m_t^2}{m_{\wt t_1}^2m_{\wt t_2}^2}\R\simeq0.4.
\end{align}
To get the estimation we have used Eq.~(\ref{inputs}) and set $\tan\beta=2,$ $\mu=200$ GeV. On the other hand, stops
decrease the width of Higgs decay to di-photon. The degree of decrease is about 20$\%$, if the reduced coupling
ratio between Higgs and vector bosons is quite close to 1. So the light stop sector only mildly increase the di-photon
rate by $\sim10\%$. Generically, the excess from light stops is about a few times $10\%$'s. But it is still very important
to isolate this contribution, and we shall leave it for our future work.
  \item By contrast, it is harder to achieve a relatively heavy Higgs boson and obvious di-photon excess in the pulling
 scenario. In fact, the excess tends to be slightly below the SM prediction.
\end{itemize}

\begin{figure}[htb]
\begin{center}
\includegraphics[width=3.21in]{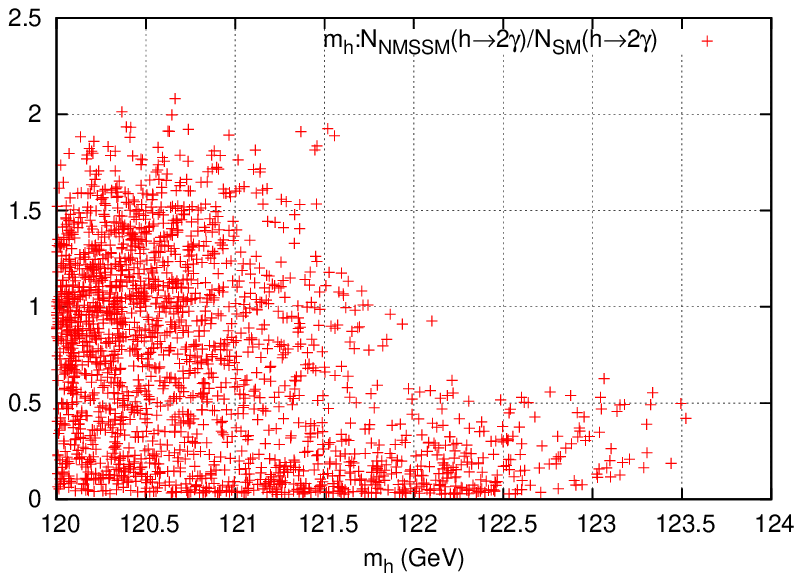}
\includegraphics[width=3.21in]{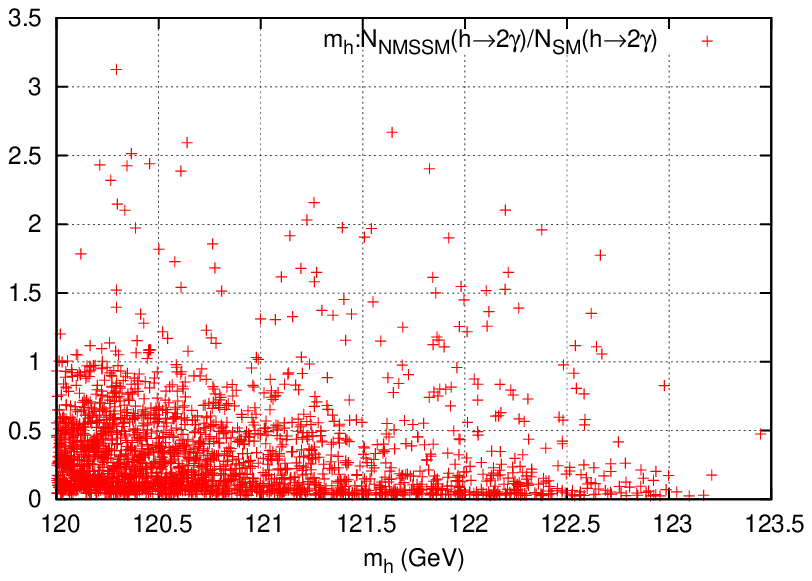}
\end{center}
\caption{\label{hgamma:2} Di-photon  rates versus the Higgs boson mass in the small $\ld$ scenario. Left plot is
for $m_{\wt q}=400$ GeV and $A_t=800$ GeV while the right for $m_{\wt q}=450$ GeV and $A_t=1100$ GeV.}
\end{figure}

In the small $\ld$ scenario, $m_h$ is expected to be pushed totally by the mixing effect.
Although here vacuum stability does not recur fine-tuning, we suffer a severer
dependence of $m_h$ on the stop sector. Consequently, the fine-tuning is expected to be worse. Additionally,
the $H_u^0$ component in $h$ is considerably reduced in most cases as discussed in Section~\ref{h:smallld}. So
the diphoton excess is not so spectacular as in the large $\ld$ scenario. Fig.~\ref{hgamma:2} supports the conclusion:
We have only scanned the regions near the most optimistic points, but the points showing
significant di-photon excess are still rare.

The recent hints favor the pushing scenario with a large $\ld$, but we are looking forward to the confirmative precise
measurements of the Higgs boson properties to distinguish different scenarios.

%%%fig.2%%%%%%%%%%%%%%%%
%%%%%%%%%%%%%%%%%%%%%%%%%%%%%%%%%%%%%%%%%%%%%%%%%%%%%%%%%%%%%%%%%%%

\subsection{The neutralino LSP dark matter: a challenge from  XENON100 experiment}\label{}

\begin{figure}[htb]
\begin{center}
\includegraphics[width=3.21in]{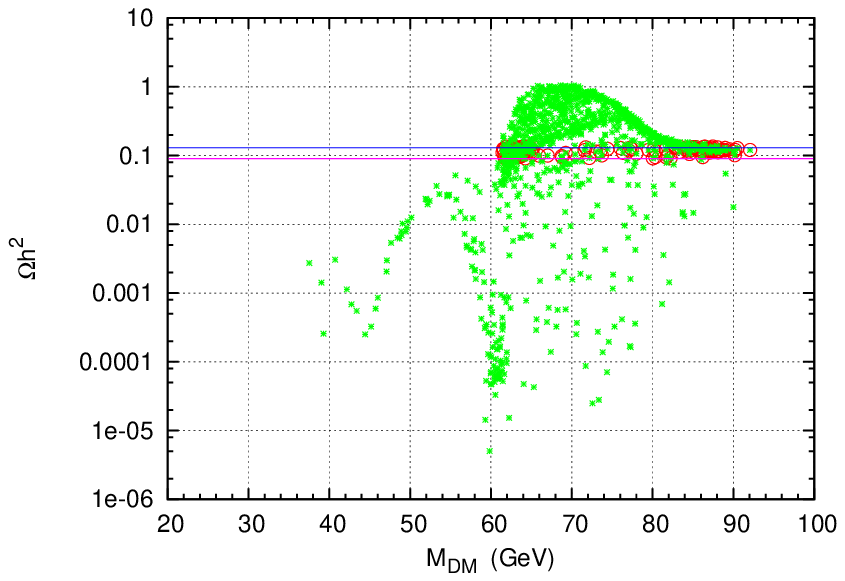}
\includegraphics[width=3.21in]{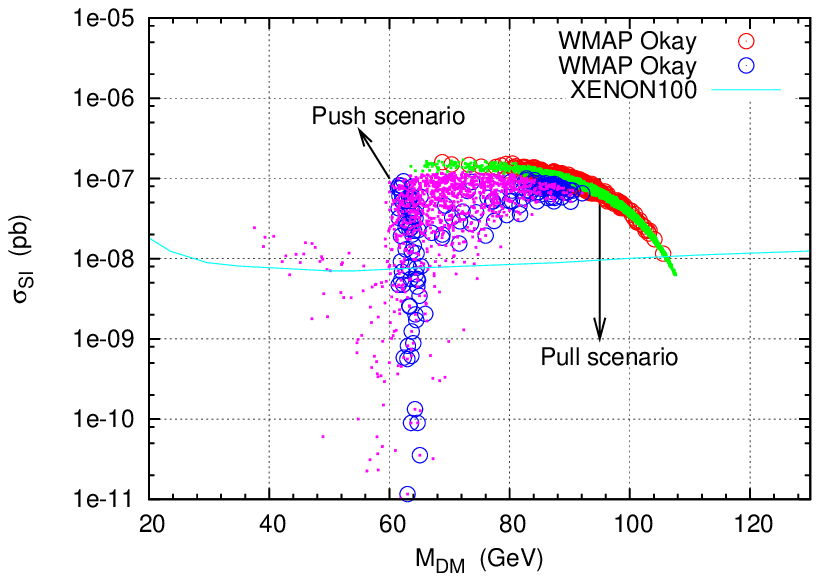}
\end{center}
\caption{\label{LSP} Left: The LSP neutralino relic density under the WMAP constraint versus its mass. The annihilation rate is not problematic due to the large Higgsino component. Right: the LSP neutralino-proton
spin-independent recoiling cross section under the XENON100 experimental constraint versus its mass.
It is strongly disfavored. Here  $\ld=0.60$.}
\end{figure}

The existence of dark matter is an important evidence for new physics, maybe regarded as a triumph for SUSY
which naturally provides the neutralino LSP as a WIMP dark matter candidate. So it is necessary to investigate
the naturalness implication on the neutralino DM. By definition, the natural NMSSM (We focus on the most
interesting case with a large $\ld$) displays a light neutralino world: the small $M_{1/2}$ and $\mu$, and
a light singlino $\wt s$ as well due to $M_{\wt s}=2\kappa v_s\lesssim\ld v_s=\mu$ in the NMSSM.
Furthermore, a large $\ld$ implies that the bino, Higgsino, and singlino are well mixed in the neutralino.
Consequently, the large Higgsino component~\footnote{Especially in the heavy $M_{\chi_1}$ region, which
is favored on account of suppressing $h$ invisible decay as well as the naturalness requiring
the larger $\kappa$.} makes the LSP strongly disfavored by the XENON100 experiment. The Higgs bosons $H_{i}$
mediate the DM-proton interaction, which gives rise to the spin-independent (SI) cross section $\sigma_{p}$.
At the zero-momentum transfer it is formulated as $\sigma_p={4\mu_p^2} f_p^2/{\pi}$~\cite{Jungman:1995df} with
the reduced mass $\mu_p=M_{\rm DM}m_p/(M_{\rm DM}+m_p)\simeq m_p$, and
\begin{align}\label{SIF}
f_p=&\f{g_{H_i\bar \chi\chi} }{m_{H_i}^2}\L\f{m_p}{v\sin\beta}\R\sum_{a=1,2,3}\left[O_{iH_u}f^{(p)}_{T_{u_a}}+ O_{iH_d}\,\tan\beta\,f^{(p)}_{T_{d_a}}\right]\cr
\approx&\L O_{iH_u}0.144+0.206O_{iH_d}\tan\beta\R \L\f{m_p}{v\sin\beta}\R \f{g_{H_i\bar \chi\chi}}{m_{H_i}^2},
\end{align}
where the first and second terms in the bracket respectively denote the up-type and down-type quark contributions.
We have used $f_{T_u}^{(p)}=0.020\pm0.004$, $f_{T_u}^{(n)}=0.014\pm0.003$, $f_{T_d}^{(p)}=0.026\pm0.005$, $f_{T_d}^{(n)}=0.036\pm0.008$,
and $f_{T_s}^{(p,n)}=0.118\pm0.062$~\cite{NF}. Because the up-type quarks give the dominant contributions, we get
\begin{align}\label{uplimit}
f_p
\simeq&1.5\times10^{-8}\times \L\f{0.9}{\sin\beta}\R \L\f{g_{H_i\bar \chi\chi}}{0.3}\R\L\f{O_{iH_u}}{0.9}\R\L\f{125\rm\,GeV}{m_{H_1}}\R^2.
\end{align}
On the other hand, for a typical WIMP with mass $\sim20-100$ GeV, the XENON100 experiment
upper bound is $f_p\simeq 0.5\times 10^{-8}$ GeV$^{-2}$. So generically $\sigma_p$ is
about one order larger than the experimentally allowed bound, justified by Fig.~\ref{LSP}.

%%%fig.2%%%%%%%%%%%%%%%%
%%%%%%%%%%%%%%%%%%%%%%%%%%%%%%%%%%%%%%%%%%%%%%%%%%%%%%%%%%%%%%%%%%%

\begin{figure}[htb]
\begin{center}
\includegraphics[width=3.21in]{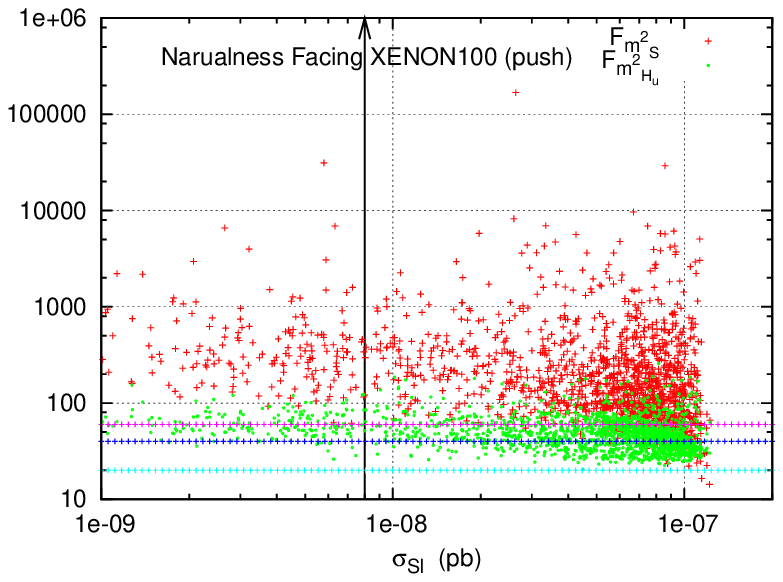}
\includegraphics[width=3.21in]{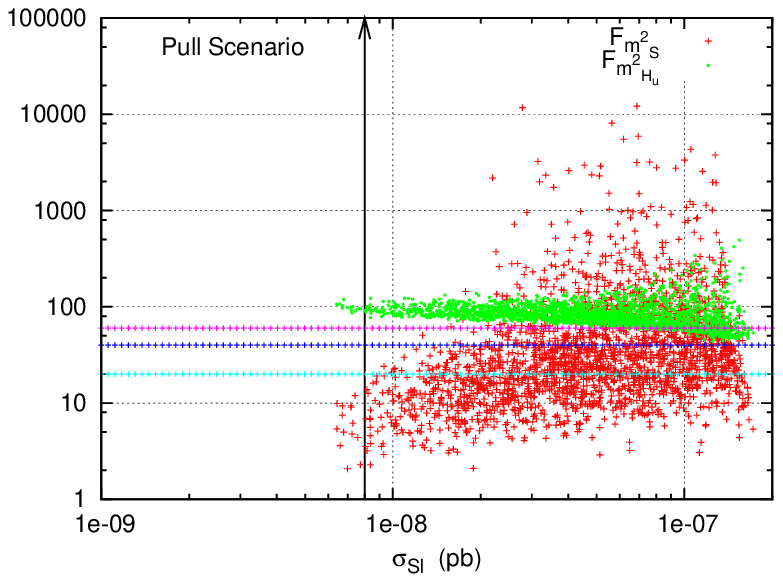}
\end{center}
\caption{\label{FLSP} The naturalness implications on the   LSP neutralino.
 The arrow indicates the XENON100 experimental bound. Here, we have fixed $\ld=0.60$.}
\end{figure}

However, the LSP can still circumvent the exclusion in a subtle way. The point is that the coupling of the
vertex $g_{211}H_2\chi_1\chi_1$ can be suppressed considerably as a result of accidental cancellation
(We noticed that this point has also been mentioned in Ref~\cite{Cerdeno:2004xw}.). Interestingly,
in the pushing scenario the WMAP and XENON100 experiments together ``predict" a LSP with mass in the
vicinity of $m_h/2\simeq63$ GeV. While in the pulling scenario the cancellation happens for the heavier
neutralino, so it ``predicted" a larger  mass $\sim100$ GeV, as shown in the right panel of Fig.~\ref{LSP}.
If we relax the naive WMAP bound via a non-standard thermal history of the Universe, the XENON100
experiment leaves more room for the neutralino LSP in the pushing scenario, as shown in Fig.~\ref{LSP}.
However, once the naturalness bound is further imposed, this case is again not favored. Fig.~\ref{FLSP} shows that
in the region allowed by the XENON100 experiment the degree of fine-tuning is typically below $1\%$,
in both the pulling and pushing scenarios. Therefore, we may want to consider the other more natural
LSP DM candidates, such as the sneutraino~\cite{sneutrino}.

Some remarks are in orders. (A) For a small $\ld$, the neutralino LSP passing the constraints also possess
a large Higgsino component. But this scenario accommodates no cancellation, and then hardly reconciles with
the XENON100 experimental bound. (B) A few works~\footnote{For the general LHC bound on the DM via the monojet search,
please see Ref.~\cite{collider}.} have studied the correlation between the Higgs and DM searchs as well~\cite{Cao:2011re}.
However, in the NMSSM a relatively heavy Higgs boson, following naturalness criteria, has more direct impact on the neutralino LSP DM:
It prefers the light neutralino world, which is however severely disfavoured by the XENON100 experiment.
Of course, this correlation between the naturalness and LSP DM search can be removed, if we give up the hypothesis of gaugino
mass unification. In that case the LSP can be bino-like and hence evade the constraint from the naturalness discussed here.
(C) If we stick to gaugino mass unification, \emph{the scenario with low mediation scale stands out again, where the LSP
is the gravitino which obviously escapes the XENON100 experimental bound.}.

%%%fig.2%%%%%%%%%%%%%%%%
%%%%%%%%%%%%%%%%%%%%%%%%%%%%%%%%%%%%%%%%%%%%%%%%%%%%%%%%%%%%%%%%%%%

\section{Discussion and conclusion}

We considered the naturalness of  $m_Z$ and $m_h$ in the MSSM and NMSSM via the bottom-up approach.
For the GUT-scale SUSY breaking mediation with light stops,  the gluino dominates the source of fine-tuning of $m_Z$,
and the present LHC lower bound already  gives   $\Delta_Z<2.5\%$.
Also, the relatively heavy  $m_h$ near  125 GeV  strongly favors  the NMSSM in view of naturalness.
 Then we studied the Higgs boson mass for a few scenarios in the NMSSM:
(1) A large $\ld$ in the  pushing scenario
which has a heavy Higgs boson and an significant di-photon excess.   However,
   this scenario   has a new large  fine tuning, owing  to vacuum stability and large $\ld-$RGE effects.
(2) A large $\ld$ but in the pulling scenario,     $\ld\sim0.7$ and  $\tan\beta\sim2$ allows  a heavy   $m_h$.  The
 difference between  this scenario and the above one is the absence of
a lighter Higgs boson. And the di-photon excess may not be generated in this case.
(3) A  small $\ld$,    the mixing effect   is still able to give a heavy enough Higgs boson.
 However, it  may not have the significant di-photon excess. Also,  the
naturalness status becomes worse in both the second and third cases.
In all these scenarios the LSP neutralino DM
is strongly disfavoured by the XENON100 experiment.  Even if the LSP neutralino DM is fine, the naturalness  still disfavors it.
Note that the above analyses are based on the mSUGRA-like model with GUT-scale
mediation,  whereas for the low scale mediation the naturalness   can be improved by about one order.

There are still some open questions to realize the natural NMSSM. For example, how light the  stops that  the LHC can tolerate  is still an important  problem in the light stop framework, although some   preliminary  works appeared~\cite{Brust:2011tb}.
 We   stress that although the discussion on the new fine-tuning associated with $m_S$ is confined to the conformal NMSSM in this work,   it may applies to the general NMSSM where the cancellation discussion  in this work is  still required.

\section*{Acknowledgement}

We would like to thank U. Ellwanger, Yunjie Huo, and Zheng Sun very much for helpful discussions,
 and thank Tao Liu very much for the collaboration in the early stage of this project.
This research was supported in part
by the Natural Science Foundation of China
under grant numbers 10821504, 11075194, 11135003, and 11275246,
and by the DOE grant DE-FG03-95-Er-40917 (TL).

\appendix

\section{An approximated  solution to the Higgs boson masses}\label{CPeven:Higgs}

  For further convenience,  we present the neutral Higgs fields in a basis with explicit Goldstone boson $G^0$~\cite{PQ:mass},
\begin{align}\label{}
H_u^0=&v_u+\f{1}{\sqrt{2}}(S_1\cos\beta+S_2\sin\beta)+\f{i}{\sqrt{2}}(P_1\cos\beta+G^0\sin\beta),\cr
H_d^0=&v_d+\f{1}{\sqrt{2}}(-S_1\sin\beta+S_2\cos\beta)+\f{i}{\sqrt{2}}(P_1\sin\beta-G^0\cos\beta),\cr
S=&v_s+\f{P_2+iS_3}{\sqrt{2}}.
\end{align}
  Thus, the elements of the CP-even Higgs boson mass squared matrix $M_S^2$ (in the basis $(S_1,S_2,S_3)$) are given by
\begin{align}\label{Higgs:even}
&(M_S^2)_{11}=M_A^2+(m_Z^2-\ld^2v^2)\sin^22\beta,\cr
& (M_S^2)_{12}=-\f{1}{2}(m_Z^2-\ld^2v^2)\sin4\beta,\cr
&(M_S^2)_{13}=-(M_A^2\sin2\beta+2\ld\kappa v_s^2)\cos2\beta\f{v}{v_s},\cr
&(M_S^2)_{22}=m_Z^2\cos^22\beta+\ld^2v^2\sin^22\beta,\cr
&(M_S^2)_{23}=\f{1}{2}(4\ld^2v_s^2-M_A^2\sin^22\beta-2\ld\kappa v_s^2\sin2\beta)\f{v}{v_s},\cr
&(M_S^2)_{33}=\f{1}{4}M_A^2\sin^22\beta\L\f{v}{v_s}\R^2+4\kappa^2v_s^2+\kappa A_\kappa v_s-\f{1}{2}\ld\kappa v^2\sin2\beta,
\end{align}
where  $M_A^2=2\ld v_s(A_\ld+\kappa v_s)/\sin2\beta$.
The doublet block has been approximately  diagonalized  already, with  $(M_S)^2_{22}$  being the frequently  quoted tree-level upper bound on the
CP-even Higgs boson mass square. For completeness, we also give  the two CP-odd Higgs boson mass squared matrix  elements:
\begin{align}\label{Higgs:odd}
&(M_P^2)_{11}=M_A^2,\cr
& (M_P^2)_{12}=\f{1}{2}(M_A^2\sin2\beta-6\ld\kappa v_s^2)\f{v}{v_s},\cr
&(M_S^2)_{22}=\f{1}{2}(M_A^2\sin2\beta+6\ld\kappa v_s^2)\sin2\beta\,\L\f{v}{v_s}\R^2-3\kappa A_\kappa v_s.
\end{align}

Considering   the 3$\times3$ CP-even Higgs boson mass squared matrix structure, we can find
the quite precise approximate eigenvalues analytically.
 First, we  diagonalize the 13-block and get two eigenvalues:
\begin{align}
m_3^2\approx M_{11}^2+(M_{S}^2)_{13}^4/M_{11}^2,\quad m_{1'}^{2}\approx M_{22}^2-(M_{S}^2)_{13}^4/M_{11}^2,
\end{align}
with the corresponding eigenvectors dominated by the first component.
 Now the heaviest eigenvalue  is isolated, and the interesting  two lighter eigenvalues can be obtained from the following effective 23-block:
\begin{align}
{\cal M}^2=\L\begin{array}{cc}
               M_{22}^2 & M_{23}^2 \\
               M_{23}^2 & m_{1'}^{2}
             \end{array}\R.
\end{align}
Thus, it can be solved analytically.

\section{The $m_S^2$ dependence of $\kappa$}\label{kappa:dep}

\begin{figure}[htb]
\begin{center}
\includegraphics[width=3.2in]{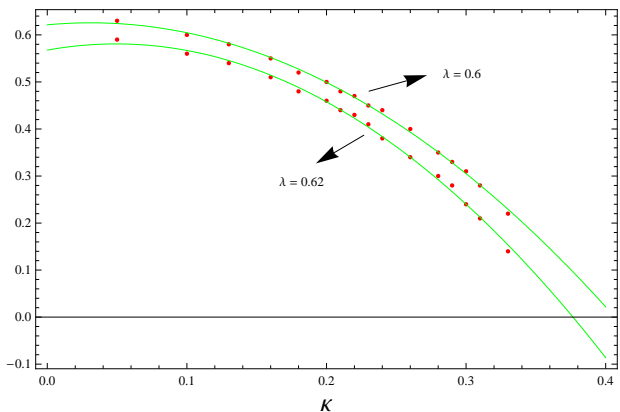}
\end{center}
\caption{\label{FIT} The  $\bar m_S^2$ component in   $m_S^2$ versus   $\kappa$ for  $\ld=0.60$ and $\ld=0.62$.}
\end{figure}

 $m_S^2$ is sensitive to  $\kappa$  in the large $\ld$ limit.  The  fraction of the $\bar m_S^2$ component
in $m_S^2$, denoted as $C_{m_S^2}(\kappa,\ld)$,   is important for the fine-tuning calculations.
  For $\ld=0.62$, we use the following  function to fit $C_{m_S^2}(\kappa,\ld)$:
 \begin{align}\label{}
C_{m_S^2}(\kappa,\ld=0.62)\approx 0.57+ 0.54\kappa - 5.42\kappa^2.
  \end{align}
And for   $\ld=0.60$, the fitting  function is
 \begin{align}\label{}
C_{m_S^2}(\kappa,\ld=0.60)\approx 0.62+ 0.28\kappa - 4.43\kappa^2.
  \end{align}
Moreover, the fitting function  plot  is given in Fig.~\ref{FIT}.

%%%fig.2%%%%%%%%%%%%%%%%
%%%%%%%%%%%%%%%%%%%%%%%%%%%%%%%%%%%%%%%%%%%%%%%%%%%%%%%%%%%%%%%%%%%

\end{document}